\documentclass[fleqn,usenatbib,onecolumn]{mnras}

\usepackage[T1]{fontenc}
\usepackage{ae,aecompl}
\usepackage{longtable}

\usepackage{graphicx}	
\usepackage{amsmath}	
\usepackage{amssymb}	
\usepackage{newtxtext,newtxmath}

\def\nh{{N$_{\rm H}$}}
\def\cm2{{cm$^{-2}$}}

\title[Galaxy pairs in 3XMM]{An XMM-Newton study of active-inactive galaxy pairs}

\author[M. Guainazzi et al.]{Matteo Guainazzi$^{1},$\thanks{E-mail: Matteo.Guainazzi@sciops.esa.int}
Alessandra De Rosa$^{2}$,
Stefano Bianchi$^{3}$,
Bernd Husemann$^{4}$,
\newauthor
Tamara Bogdanovic$^{5}$,
Stefanie Komossa$^{6}$,
Nora Loiseau$^{7}$,
Zsolt Paragi$^{8}$,
\newauthor
Miguel P\'erez-Torres$^{9}$,
Enrico Piconcelli$^{10}$,
Cristian Vignali$^{11}$
\\
$^{1}$ESA European Space Research and Technology Centre (ESTEC), Keplerlaan 1, 2201 AZ, Noordwijk, The Netherlands\\
$^{2}$INAF - Istituto di Astrofisica e Planetologia Spaziali (IAPS), via Fosso del Cavaliere, I-133 Roma, Italy\\
$^{3}$Dipartimento di Matematica e Fisica, Universit\`a degli Studi Roma Tre, via della Vasca Navale 84, I-00146 Roma, Italy\\
$^{4}$Max Planck Institute for Astronomy, K\"onigstuhl 17, 69117, Heidelberg, Germany \\
$^{5}$Center for Relativistic Astrophysics, School of Physics, Georgia Institute of Technology, 837 State Street, Atlanta, GA 30332-0430, United States \\
$^{6}$Max-Planck-Institut f\"ur Radioastronomie, Auf dem H\"ugel 69, D-521 Bonn, Germany \\
$^{7}$ESA - European Space Astronomy Centre (ESAC), E-28692 Villanueva de la Ca\~nada, Madrid, Spain \\
$^{8}$Joint Institute for VLBI ERIC, Oude Hoogeveensedijk 4, 7991 PD, Dwingeloo, The Netherlands \\
$^{9}$Instituto de Astrof\'isica de Andaluc\'ia (IAA-CSIC),
              Glorieta de la Astronom\'ia s/n, 18008 Granada, Spain \\
$^{10}$Osservatorio Astronomico di Roma (INAF), via Frascati 33, 00040, Monte Porzio Catone, Roma, Italy \\
$^{11}$Dipartimento di Fisica e Astronomia, Alma Mater Studiorum, Universit\`a degli Studi di Bologna, Via Gobetti 93/2, 40129, Bologna, Italy
}

\date{Accepted 2021 March 15. Received 2021 March 1; in original form 2020 December 10}

\pubyear{2020}

\begin{document}
\label{firstpage}
\pagerange{\pageref{firstpage}--\pageref{lastpage}}
\maketitle

\begin{abstract}
While theory and simulations indicate that galaxy mergers play an important role in the cosmological evolution
of accreting black holes and their host galaxies, samples of Active Galactic Nuclei (AGN) in galaxies at close
separations are still small.
In order to increase the sample of AGN pairs, we undertook an archival project to investigate
the X-ray properties of a SDSS-selected sample of
32 galaxy pairs with separations $\le$150~kpc
containing one optically-identified AGN, that were serendipitously observed by XMM-Newton. We discovered
only one X-ray counterpart among the optically classified non-active galaxies, with a weak X-ray luminosity ($\simeq$5$\times$10$^{41}$~erg~s$^{-1}$). 59\% (19 out of 32) of the AGN in our galaxy pair sample exhibit an X-ray
counterpart, covering a wide range in absorption corrected X-ray luminosity (5$\times$10$^{41}$--2$\times$10$^{44}$~erg~s$^{-1}$). More than 79\% of these AGN are
obscured (column density $N_H>$10$^{22}$~cm$^{-2}$), with more than half thereof ({\it i.e.}, about 47\% of the total AGN sample) being Compton-thick.
AGN/no-AGN pairs are therefore more frequently X-ray obscured (by a factor $\simeq$1.5)
than isolated AGN. When compared to a luminosity and redshift-matched sample of {\it bona fide}
dual AGN, AGN/no-AGN pairs exhibit one order-of-magnitude lower X-ray column density in the same separation
range ($>$10~kpc). A small sample (4 objects) of AGN/no-AGN pairs 
with sub-pc separation are all heavily obscured, driving a formal
anti-correlation between the X-ray column density and the galaxy pair separation in these systems. 
These findings suggest that the galactic environment has a key influence on the triggering
of nuclear activity in merging galaxies.
\end{abstract}




\section{Introduction}

The tight empirical correlations between the mass of
super-massive black holes, and observable quantities
related to
the host galaxy size (such as the "M-$\sigma$ relation; \citealt{ferrarese00, mcconnell13, kormendy13}) have been interpreted as signature
of the cosmological co-evolution of accreting black
holes and host galaxies. In this context, galaxy
merging could play an important role in triggering the
formation of black holes via direct gas collapse \citep{begelman06, mayer10}, generating gas instabilities
that allow the nuclear gas to lose angular momentum and
efficiently feed the black hole \citep{dimatteo05}, or
even shape the environment and clustering properties of
quasars \citep{kauffmann00, alexander12}.
However, it remains unclear if there is a relation between
galaxy interaction and nuclear activity. Results from
X-ray and IR surveys indicate an increase of the fraction
of Active Galactic Nuclei (AGN) in the population
of merging galaxies for decreasing nuclear separation
below 100~kpc \citep{ellison11, silverman11, satyapal14}. Also the fraction of "dual AGN" -
close
galaxy companions hosting both actively accreting black
holes - increases with decreasing separation, and their
X-ray luminosity increases \citep{koss12}.
However, these conclusions
are based on relatively small samples. Dual AGN are rare,
and many of the known specimen of this class have been
discovered serendipitously. A robust comparison with
models of galaxy mergers evolution is often
cumbersome \citep{derosa19}.

One of the main observational biases is the fact
the AGN pairs are often heavily dust-enshrouded.
It has been recently demonstrated that obscured
AGN show a significant excess of late-stage nuclear mergers
when compared to a control sample of inactive galaxies
\citep{koss18}.
Identifying
dual AGN via optical spectroscopy is therefore hard. Together with IR \citep{imanishi14},
X-rays are an efficient way to detect accretion-powered
processes in low-to-moderately obscured
sources ({\it i.e.} Compton-thin;
column density $N_H$~$\le$~10$^{24}$~cm$^{-2}$),
and even heavily obscured sources can be identified in
deep exposures (through optically-thick reprocessing
spectral components; see , {\it e.g.}, \citet{piconcelli10}), or through measurements above
10~keV \citep{bianchi08, koss11, iwasawa18}.
Interestingly, X-ray obscuration seems also to
increase with decreasing AGN separation \citep{ricci17, derosa18, pfeifle19}. This is in agreement with numerical
simulations \citep{capelo17}, even though they
typically probe
spatial scales ($\ge$~50~pc) much larger than the scale
of the torus where the bulk of X-ray obscuration is
assumed to occur. However, these results are
somewhat dependent
on the sample selection criteria.
Optically-selected dual AGN samples tend to show a
larger fraction of unobscured or moderately obscured AGN
than X-ray selected samples \citep{green11, gross19}.
Whether this is the correction of a bias, or a bias in
itself it is still to be ascertained.

In order to increase the size of well-defined
AGN pair samples, we undertook an archival project to investigate
the X-ray properties of a SDSS-selected sample of
galaxy pairs with separations $\le$150~kpc
containing one optically-identified AGN and serendipitously observed by XMM-Newton
\citep{jansen01}. The original goal of the study was
identifying dual AGN pairs that may have been missed
because the
nuclear activity in the optical inactive galaxy remains
undetected due to, for
instance, heavy nuclear obscuration. The results of the
study yield new observational constraints on the
trigger mechanism of supermassive black holes in merging
galaxies at separations $\le$100~kpc.

The paper is structured as follows. In \S~\ref{sect:sample} we present
the sample selection criteria. A re-analysis of the SDSS imaging
and spectroscopy data is presented in
\S~\ref{sect:optical_spectroscopy}; this is required to validate
the nature of the galaxy pairs discussed in this
paper. The X-ray data analysis
is presented in \S~\ref{sect:reduction}.
Our results are discussed in \S~\ref{sect:discussion} and the main
conclusions are summarized in \S~\ref{sect:conclusions}.

\section{Sample selection}
\label{sect:sample}

\begin{figure}
    \centering
   \includegraphics[width=0.75\textwidth]{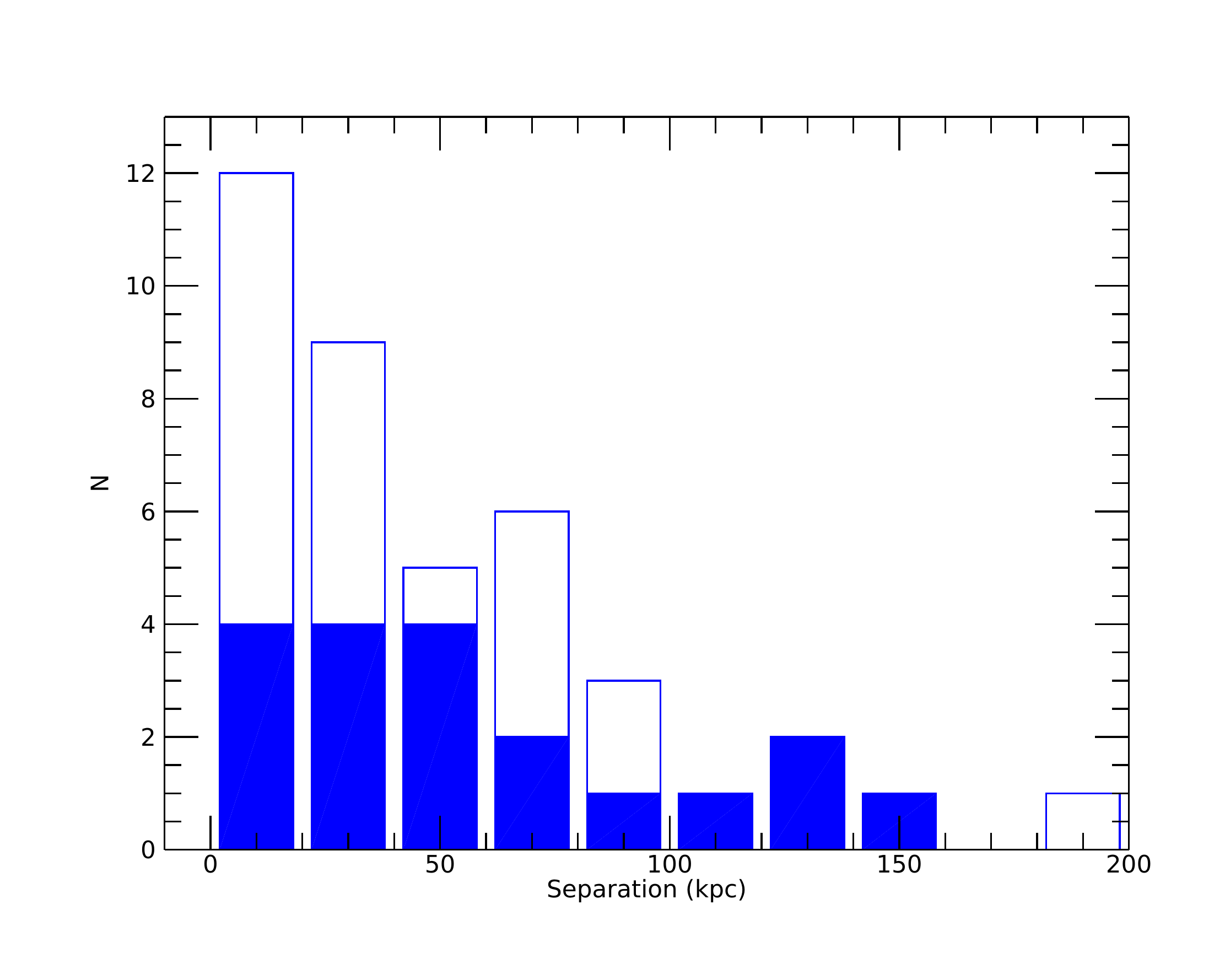}
    \caption{Histogram of the galaxy separations for the galaxies in our sample with at least one XMM-Newton observations ({\it empty bars}) and for those with at least one X-ray counterpart ({\it filled bars}) (cf. Tab.~\ref{tab_xraycount})}
    \label{fig:separations}
\end{figure}

In order to select the AGN/no-AGN pairs discussed in this paper, we took the SDSS DR7 galaxy sample from the MPA-JHU value added catalogs and classified all emission line galaxies into star forming galaxies, LINERs and Seyfert 2 galaxies using classical BPT
diagnostics \citep{baldwinetal81}. All galaxies with too weak emission lines for a BPT classification (S/N<3), were assigned the class "UNCLEAR". Furthermore, we considered 3600 unobscured AGN from the SDSS DR7 derived catalog of \citet{stern12} as Seyfert 1 galaxies. Based on the combined Seyfert 1 and Seyfert 2 catalog we identified potential companion galaxies, classified as star forming, LINERs, or UNCLEAR, with a projected
separation, $R$, of 5~kpc$<$R$<$200~kpc and a redshift difference of $\Delta$z$<0.01$.
This leads to 698 Seyfert-SF pairs, 71 Seyfert-LINER pairs and 780 Seyfert-UNCLEAR pairs. We have then searched in the archive of XMM-Newton observations all the fields where each pair was observed at least once with at least one of the EPIC cameras. This search yielded 41 matches. 9 of these sources were later discarded because a reanalysis of the SDSS imaging and spectroscopic data did not confirm the original classification (cf. \S~\ref{sect:optical_spectroscopy}). The final list includes 32 pairs, whose coordinates and redshifts are shown in Tab.~\ref{tab_ul}. The galaxy pairs with at least one X-ray
counterpart (cf. \S~\ref{sect:photometry}) cover
a range in black hole separation up to $\simeq$150~kpc,
with a median separation of $\simeq$40~kpc (Fig.~\ref{fig:separations}). All 32 sky positions were observed by
XMM-Newton only once, and only a few of them were at the boresight of the corresponding observation.

Our optical selection biases the sample against heavily-obscured (buried) AGN, whose ionizing photons cannot create a NLR. On the other hand, using X-ray spectroscopic data taken with the EPIC camera on board XMM-Newton for the spectral fitting may also prevent us from detecting objects whose X-ray emission is suppressed by heavy obscuration. Thus, the fraction of obscured AGN that we derive from our sample is likely to constitute a lower limit to that of the parent sample of galaxy pairs with at least one AGN. We discuss in \S~\ref{sect:optical_spectroscopy} a way to at least partly alleviate the latter bias.

\section{Optical spectroscopy}

\label{sect:optical_spectroscopy}
\begin{figure}
    \centering
   \includegraphics[width=\textwidth]{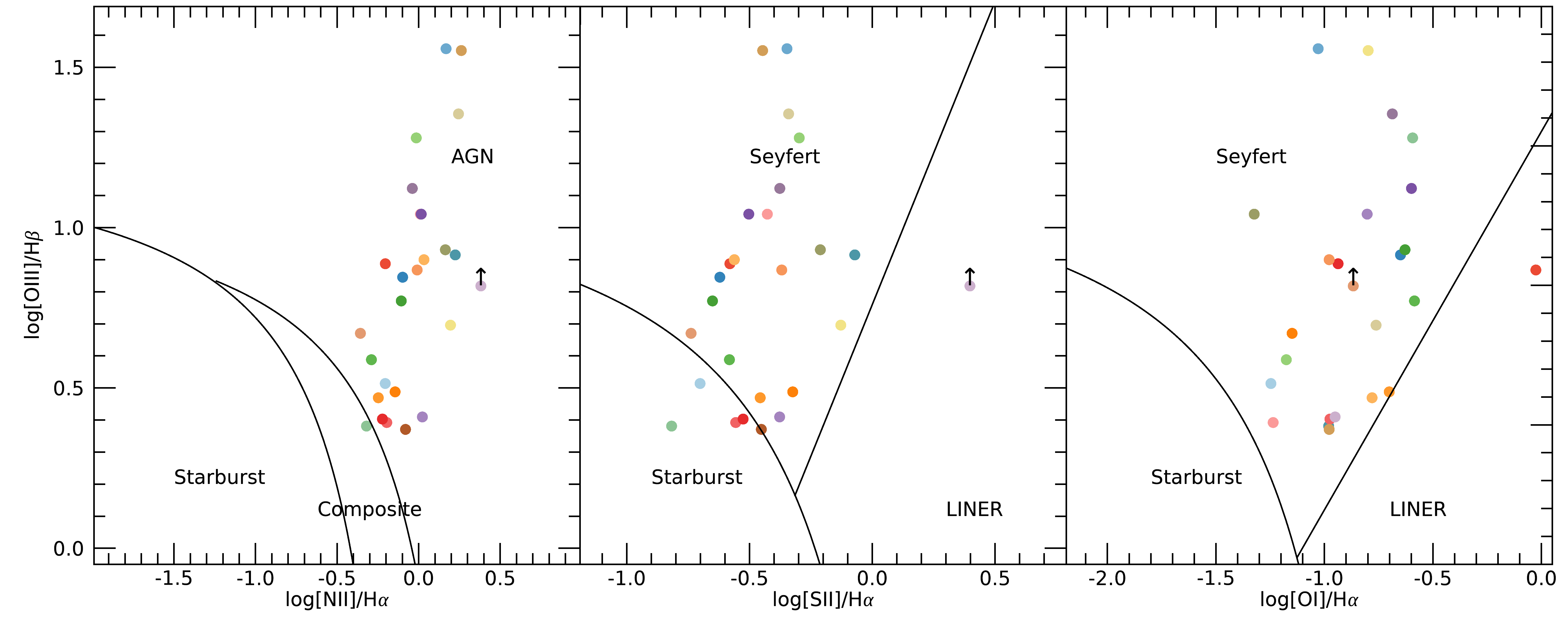}
    \caption{BPT diagrams for all the sources of our sample. See text for details.}
    \label{fig:bpt}
\end{figure}

Aware of possible inaccuracies in the SDSS spectroscopic classification of close galaxy pairs due to the spillover of the primary flux to the secondary fibre \citep{husemann20}, we retrieved the SDSS DR12 spectra \citep{SDSS-DR12} at the location of the galaxies, as reported in Table \ref{tab_ul}, from the survey webpage\footnote{http://skyserver.sdss3.org/dr12.}. 
We analyse the full-band optical spectra with the software package QSFit 1.3.0 \citep{qsfit}, which automatically takes into account both the AGN and the host galaxy emission, together with a number of broad and narrow emission lines. In particular, we extracted the intensities of the the primary diagnostic narrow emission lines such as H$\beta$, [OIII] $\lambda5007$, [OI] $\lambda6300$, H$\alpha$, [NII] $\lambda6583$ and [SII] $\lambda\lambda6717,6730$. Five sources (namely SSDSSJ002920.35-001028.9, SDSSJ011254.91+00031.0, SDSSJ011448.67-002946.0, SDSSJ111830.28+402554.0, SDSSJ170601.86+601732.3) show unambiguously the presence of dominant broad components of the Balmer lines, and are therefore included in our sample. For all the other sources, the line flux ratios were then plotted in the BPT diagrams shown in Fig.~\ref{fig:bpt}, where the separations between Seyfert galaxies, starburst galaxies, and low-ionization nuclear emission-line regions (LINERs) are as in \citet{kewley2006}. Sources with at least one BPT diagram indicating an AGN are kept in our sample, while LINERs and all the other cases are excluded.

The ratio between the observed 2--10~keV and [OIII]$\lambda5007$ luminosity in AGN is often used as a proxy of X-ray obscuration  \citep{bassani99}, that is one of the objectives of our study. 
The X-ray luminosity of our sample will be discussed in the sect. \ref{sect:spectroscopy}, while to derive the L$_{\rm [OIII]}$ corrected for extinction, we assumed the \cite{cardelli89} extinction law and an intrinsic Balmer decrement (H$\alpha$/H$\beta$) equal to 3, which represents the case for the Narrow Line Region (NLR) \citep{osterbrock06}.
We then will use the ratio between the X-ray and ${\rm [OIII]}$ $\lambda$5007 luminosity as an indirect measurement of \nh, to be compared with the value obtained through X-ray spectroscopy (see Sect.\ref{sect:spectroscopy}).

\section{X-ray data analysis}
\label{sect:reduction}

For each of the XMM-Newton observations we
extracted the Observation Data Files (ODF) from the
XMM-Newton Science Archive (XSA). The list of
Observation Identifiers (Obs.\#) is shown
in Tab.~\ref{tab_ul}. The data of the
EPIC cameras \citep{strueder01,turner01} were reduced using SAS version 16.0 \citep{gabriel04}, and the most updated calibration files available at the time the reduction was performed. Calibrated event lists were generated with the data reduction meta-tasks {\tt epproc} and {\tt emproc} for the EPIC-pn and the EPIC-MOS cameras, respectively.
Intervals of high particle background were removed by applying a standard threshold on a representative curve of this background component, extracted using single events with energies larger than 10~keV (10--12~keV for EPIC-pn). The value of the thresholds were 0.4 and 0.35 counts per second in EPIC-pn and EPIC-MOS, respectively.
\begin{longtable}[c]{lcccccc}
\caption{Galaxy pair sample. The "Minimum separation" indicates the separation (in arcminutes and kpc) between the AGN and the closest galaxy in the pair. The "..." sign indicates that the values of the corresponding AGN in the system shall be applied.}
\\
Obs.\# & SDSS name & RA & Dec. & z & \multicolumn{2}{c}{Minimum separation} \\
 & & & & & (") & (kpc) \\
\hline
04060101 & SDSSJ002920.36-00102 &   7.33482 &  -0.17469 & 0.061 &  43.6 &  59.7 \\
       ... &                  N/D &   7.34739 &  -0.16365 & 0.060 & ... & ... \\
       ... &                  N/D &   7.32722 &  -0.18411 & 0.060 & ... & ... \\
       ... &                  N/D &   7.34890 &  -0.17187 & 0.060 & ... & ... \\
0747390139 & SDSSJ010951.34+00024 &  17.46393 &   0.04618 & 0.086 &  54.6 & 107.2 \\
       ... &                  N/D &  17.45529 &   0.03373 & 0.090 & ... & ... \\
0747400142 & SDSSJ011254.92+000 &  18.22883 &   0.05361 & 0.239 &  23.7 & 142.1 \\
       ... &                  N/D &  18.22787 &   0.04709 & 0.240 & ... & ... \\
0404410201 & SDSSJ011429.87+00125 &  18.62445 &   0.21521 & 0.046 &  38.6 &  39.4 \\
       ... &                  N/D &  18.62415 &   0.20448 & 0.050 & ... & ... \\
0747400132 & SDSSJ011448.68-00294 &  18.70279 &  -0.49611 & 0.034 &   8.4 &   6.3 \\
       ... &                  N/D &  18.70754 &  -0.49558 & 0.030 & ... & ... \\
0747400150 & SDSSJ011659.07+00193 &  19.24612 &   0.32593 & 0.078 &  10.1 &  17.9 \\
       ... &                  N/D &  19.24880 &   0.32673 & 0.080 & ... & ... \\
0747430138 & SDSSJ014402.61-00070 &  26.01089 &  -0.11915 & 0.079 &  55.0 &  98.7 \\
       ... &                  N/D &  26.02399 &  -0.12703 & 0.080 & ... & ... \\
0142610101 & SDSSJ030639.57+00034 &  46.66483 &   0.06197 & 0.107 &  56.5 & 139.9 \\
       ... &                  N/D &  46.68293 &   0.06610 & 0.110 & ... & ... \\
       ... &                  N/D &  46.67869 &   0.06935 & 0.110 & ... & ... \\
0201120101 & SDSSJ030655.63-00014 &  46.781 &  -0.02805 & 0.112 &  50.4 & 1.1 \\
       ... &                  N/D &  46.730 &  -0.01412 & 0.110 & ... & ... \\
       ... &                  N/D &  46.72687 &  -0.04465 & 0.110 & ... & ... \\
0206340101 & SDSSJ0852.70+16261 & 133.30290 &  16.43763 & 0.065 &   6.4 &   9.4 \\
       ... &                  N/D & 133.30148 &  16.43875 & 0.060 & ... & ... \\
0725300159 & SDSSJ090255.53+01464 & 135.735 &   1.77980 & 0.118 &  22.4 &  61.6 \\
       ... &                  N/D & 135.73344 &   1.77393 & 0.120 & ... & ... \\
06750401 & SDSSJ091507.48+29562 & 138.78117 &  29.94010 & 0.1 &  23.0 &  70.8 \\
       ... &                  N/D & 138.75716 &  29.93708 & 0.130 & ... & ... \\
       ... &                  N/D & 138.77386 &  29.939 & 0.130 & ... & ... \\
0306050201 & SDSSJ094046.29+03393 & 145.19287 &   3.65839 & 0.087 &  40.4 &  80.3 \\
       ... &                  N/D & 145.18353 &   3.66466 & 0.090 & ... & ... \\
0504101701 & SDSSJ101858.47+37180 & 154.74361 &  37.30021 & 0.048 &  .0 &  33.1 \\
       ... &                  N/D & 154.73328 &  37.28583 & 0.050 & ... & ... \\
       ... &                  N/D & 154.74774 &  37.30816 & 0.050 & ... & ... \\
0556211401 & SDSSJ104856.96+59282 & 162.23734 &  59.47383 & 0.093 &  34.9 &  74.4 \\
       ... &                  N/D & 162.25620 &  59.47245 & 0.090 & ... & ... \\
0111290301 & SDSSJ111830.29+40255 & 169.62616 &  40.467 & 0.155 &   5.5 &  20.4 \\
       ... &                  N/D & 169.62616 &  40.467 & 0.160 & ... & ... \\
0047540601 & SDSSJ114713.50+47325 & 176.80624 &  47.54950 & 0.074 &  52.5 &  87.9 \\
       ... &                  N/D & 176.79112 &  47.55992 & 0.070 & ... & ... \\
0744040301 & SDSSJ115852.20+42432 & 179.71750 &  42.72242 & 0.002 &  16.4 &   0.7 \\
       ... &                  N/D & 179.64188 &  42.73404 & 0.000 & ... & ... \\
0601780601 & SDSSJ120443.32+103 & 181.18051 &  .17729 & 0.025 &  59.7 &  32.6 \\
       ... &                  N/D & 181.18834 &  .19248 & 0.030 & ... & ... \\
       ... &                  N/D & 181.18861 &  .15791 & 0.030 & ... & ... \\
       ... &                  N/D & 181.18861 &  .15772 & 0.030 & ... & ... \\
       ... &                  N/D & 181.18834 &  .19248 & 0.030 & ... & ... \\
0601780901 & SDSSJ121044.28+38201 & 182.68446 &  38.33617 & 0.023 &  50.9 &  25.5 \\
       ... &                  N/D & 182.69930 &  38.34421 & 0.020 & ... & ... \\
0722670201 & SDSSJ122846.68+07275 & 187.19452 &   7.46599 & 0.085 &   8.0 &  15.5 \\
       ... &                  N/D & 187.19670 &   7.46651 & 0.080 & ... & ... \\
0202180301 & SDSSJ124210.61+370 & 190.54420 &  33.28405 & 0.044 &  55.1 &  53.7 \\
       ... &                  N/D & 190.52699 &  33.27884 & 0.040 & ... & ... \\
0691610301 & SDSSJ125729.99+28111 & 194.42943 &  27.60787 & 0.068 &  48.7 &  74.6 \\
       ... &                  N/D & 194.41422 &  27.60609 & 0.070 & ... & ... \\
       ... &                  N/D & 194.42854 &  27.62137 & 0.070 & ... & ... \\
0055990501 & SDSSJ133817.28+48163 & 204.57198 &  48.27563 & 0.028 &  10.1 &   6.2 \\
       ... &                  N/D & 204.57404 &  48.27807 & 0.030 & ... & ... \\
0671150501 & SDSSJ134736.41+17340 & 206.90169 &  17.56796 & 0.045 &  10.0 &  10.0 \\
       ... &                  N/D & 206.90462 &  17.56780 & 0.050 & ... & ... \\
0164560701 & None                 & 211.496 &  54.41608 & 0.083 &  54.6 & 103.2 \\
       ... &                  N/D & 211.30116 &  54.42896 & 0.080 & ... & ... \\
0091140201 & SDSSJ1451.76+16552 & 223.29903 &  16.92345 & 0.045 &  49.5 &  49.4 \\
       ... &                  N/D & 223.30621 &  16.93537 & 0.050 & ... & ... \\
0721820201 & SDSSJ145840.74+38273 & 224.66077 &  38.45776 & 0.136 &  25.7 &  82.4 \\
       ... &                  N/D & 224.66972 &  38.45910 & 0.140 & ... & ... \\
0150680101 & SDSSJ151811.57+172 & 229.54819 &  .28951 & 0.108 &  30.4 &  76.0 \\
       ... &                  N/D & 229.53943 &  .29341 & 0.110 & ... & ... \\
0147210301 & SDSSJ160515.86+17422 & 241.604 &  17.70761 & 0.0 &  59.7 &  40.6 \\
       ... &                  N/D & 241.442 &  17.72411 & 0.030 & ... & ... \\
0305750301 & SDSSJ170601.87+60173 & 256.50778 &  60.29233 & 0.130 &  13.5 &  41.2 \\
       ... &                  N/D & 256.50061 &  60.29357 & 0.130 & ... & ... \\
0744410801 & SDSSJ171715.74+64154 & 259.558 &  64.26115 & 0.035 &  33.5 &  25.8 \\
       ... &                  N/D & 259.32224 &  64.27000 & 0.030 & ... & ... \\
0673000147 & SDSSJ221839.93-00240 & 334.66635 &  -0.40052 & 0.095 &  30.8 &  67.2 \\
       ... &                  N/D & 334.67072 &  -0.398 & 0.090 & ... & ... \\
\hline \hline
\label{tab_ul} \\
\end{longtable}

\subsection{XMM-Newton source detection and photometry}
\label{sect:photometry}

The first step of our analysis aimed at identifying the X-ray counterparts of the sample sources. For each optically identified pair member we run the standard source detection algorithm in the SAS, through the meta-task {\tt edetect\_chain}. In a nutshell, this task employs the detection algorithm in \citet{cruddace88} on an image whose background is determined through a 2-D spline of field-of-view map from which sources had been removed on the basis of a prior detection run.  We run the detection algorithm simultaneously on EPIC-pn and EPIC-MOS images in the 0.3--10~keV energy band extracted on a square box of 2' size around the nominal coordinates of the optically-identified AGN. We used a {\tt edetect\_chain} option allowing to simultaneously fitting the telescope Point Spread Function (PSF) in the so-called "multi-source mode" (with input parameters {\tt nmulsou}~=~2 and {\tt scut}~=~200), where neighbouring sources with overlapping PSFs are fitted simultaneously. Our choice is driven by the fact that the distance between the components of some of the pairs in our sample is comparable with the telescope angular resolution. Following \cite{rosen16}, we assume that an X-ray source is a {\it bona fide} counterpart of an optical galaxy if the centroid distance is $\le$3.5".

We show the results of our X-ray photometric analysis in Tab.~\ref{tab_xraycount}.
For each optically-identified AGN we list one row for each of the potentially associated AGN/non-AGN pair members. We significantly detect 19 X-ray sources among the optically identified AGN. For completeness, in the other cases we show the distance of the closest X-ray source in the 2' side detection box, if any.
\begin{table}
\caption{Total 0.3--10~keV count rate of the X-ray counterparts of the {\it bona fide} optically identified AGN in our sample, as estimated from the combined EPIC images of the corresponding XMM-Newton observation. Rows in {\bf bold} indicate AGN with an identified {\it bona fide} X-ray counterpart. An upper limit at the optical position (1$\sigma$-level) is shown if the closest X-ray source is located at a distance larger than the astrometric error of the XMM-Newton EPIC serendipitous source catalogue (3.5" at the 90\% confidence level). The "Distance" is between the AGN optical position and the closest X-ray source in the 2$\times$2 acrminutes$^2$ detection box. The dots indicates AGN optical positions for which no meaningful X-ray count rate upper limit could be estimated and/or no X-ray source was detected in the detection box.}
\label{tab_xraycount}
\begin{tabular}{lccc}
\hline
Source ID & Rate & Net counts & Distance \\
 & (s$^{-1}$) & & (") \\
\hline
SDSSJ002920.36-00102 & {\bf 0.688$\pm$0.040} &   {\bf  2533$\pm$125} & {\bf   0.1$\pm$  2.9} \\
SDSSJ010951.34+00024 & {\bf 0.115$\pm$0.017} &    {\bf   222$\pm$} & {\bf   5.8$\pm$  4.3} \\
SDSSJ011254.92+000 & {\bf 0.630$\pm$0.017} &    {\bf  1477$\pm$40} & {\bf   0.7$\pm$  0.2} \\
SDSSJ011429.87+00125 &            $\le$0.023 &        6632$\pm$  199 &        12.8$\pm$  1.2 \\
SDSSJ011448.68-00294 &                   ... &                   ... &                   ... \\
SDSSJ011659.07+00193 & {\bf 0.152$\pm$0.014} &    {\bf   326$\pm$28} & {\bf   6.1$\pm$  3.4} \\
SDSSJ014402.61-00070 & {\bf 0.042$\pm$0.005} &    {\bf   110$\pm$12} & {\bf   1.1$\pm$  1.0} \\
SDSSJ030639.57+00034 & {\bf 0.948$\pm$0.005} &   {\bf 33992$\pm$189} & {\bf   0.9$\pm$  0.0} \\
SDSSJ030655.63-00014 & {\bf 0.729$\pm$0.046} &   {\bf  4564$\pm$286} & {\bf   2.9$\pm$  2.8} \\
SDSSJ0852.70+16261 & {\bf 0.223$\pm$0.008} &    {\bf  2332$\pm$73} & {\bf   1.3$\pm$  0.8} \\
SDSSJ090255.53+01464 &            $\le$0.039 &          13$\pm$    5 &        77.4$\pm$  7.2 \\
SDSSJ091507.48+29562 &            $\le$0.022 &        1604$\pm$  100 &        19.6$\pm$  2.7 \\
SDSSJ094046.29+03393 &            $\le$0.019 &        2097$\pm$   93 &        56.5$\pm$  0.8 \\
SDSSJ101858.47+37180 & {\bf 0.522$\pm$0.023} &   {\bf  2379$\pm$103} & {\bf   0.6$\pm$  0.9} \\
SDSSJ104856.96+59282 & {\bf 0.289$\pm$0.009} &    {\bf  1129$\pm$36} & {\bf   1.4$\pm$  0.3} \\
SDSSJ111830.29+40255 &            $\le$2.553 &        6046$\pm$  343 &        56.3$\pm$  0.0 \\
SDSSJ114713.50+47325 &            $\le$0.014 &         762$\pm$  259 &        69.0$\pm$  1.4 \\
SDSSJ115852.20+42432 & {\bf 0.1$\pm$0.005} &    {\bf  1249$\pm$43} & {\bf   2.1$\pm$  0.7} \\
SDSSJ120443.32+103 & {\bf 0.648$\pm$0.006} &   {\bf 15460$\pm$129} & {\bf   0.5$\pm$  0.1} \\
SDSSJ121044.28+38201 & {\bf 0.601$\pm$0.040} &    {\bf  1858$\pm$47} & {\bf   1.3$\pm$  0.2} \\
SDSSJ122846.68+07275 &                   ... &                   ... &                   ... \\
SDSSJ124210.61+370 & {\bf 7.121$\pm$0.028} &   {\bf 66480$\pm$258} & {\bf   0.6$\pm$  0.0} \\
SDSSJ125729.99+28111 & {\bf 0.433$\pm$0.011} &   {\bf  5324$\pm$126} & {\bf   3.6$\pm$  1.1} \\
SDSSJ133817.28+48163 &            $\le$0.266 &        3911$\pm$  270 &        19.6$\pm$  0.2 \\
SDSSJ134736.41+17340 & {\bf 0.145$\pm$0.005} &    {\bf   887$\pm$33} & {\bf   1.6$\pm$  0.3} \\
SDSSJ1451.76+16552 &            $\le$0.022 &        3237$\pm$  262 &         9.0$\pm$  1.3 \\
SDSSJ145840.74+38273 & {\bf 0.356$\pm$0.015} &   {\bf  3681$\pm$155} & {\bf   3.5$\pm$  2.6} \\
SDSSJ151811.57+172 &            $\le$0.021 &        3652$\pm$  170 &        71.6$\pm$  1.7 \\
SDSSJ160515.86+17422 &       0.711$\pm$0.026 &        4360$\pm$  146 &         5.2$\pm$  0.7 \\
SDSSJ170601.87+60173 & {\bf 0.241$\pm$0.056} &     {\bf    48$\pm$7} & {\bf   1.4$\pm$  1.2} \\
SDSSJ171715.74+64154 & {\bf 0.004$\pm$0.002} &    {\bf    52$\pm$13} & {\bf   4.0$\pm$  0.7} \\
SDSSJ221839.93-00240 &            $\le$0.025 &            $\le$   41 &         8.5$\pm$  1.4 \\
\hline \hline
\end{tabular}
\end{table}

\begin{figure}
    \centering
    \includegraphics[width=0.5\textwidth, angle=-90]{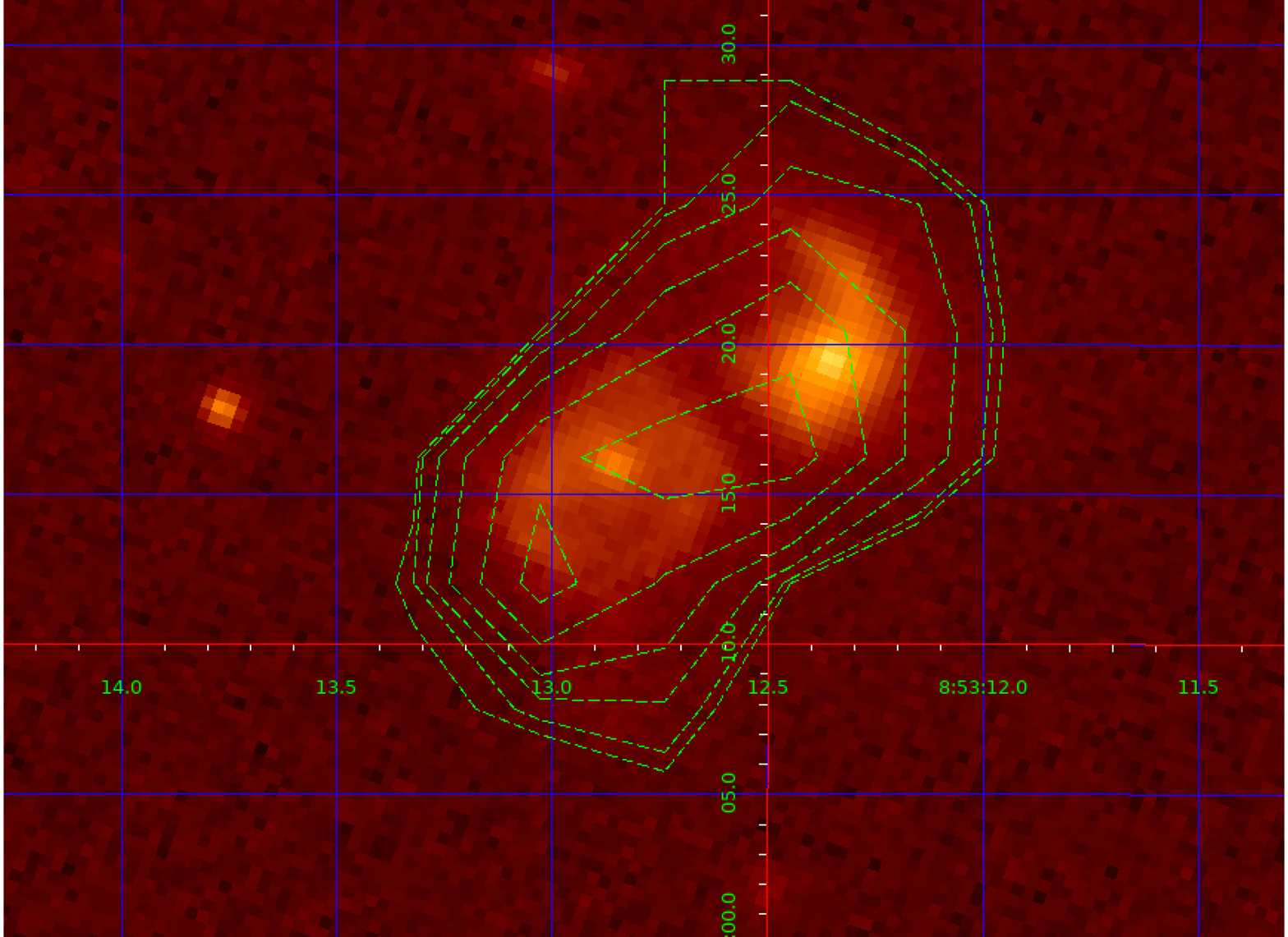}
    \caption{SDSS i-filter image superposed to the count contours of the EPIC image ({\it green dashed lines}; 6 levels in logarithmic scale in the 10--42 net EPIC counts per pixel range) for the only galaxy pair where an X-ray counterpart spatially coincident with each member of the pair is detected: SDSS~J0852.70+16261.}
    \label{fig:EPIC_vs_SDSS}
\end{figure}

Among the galaxies optically classified as "non-AGN", only one X-ray source is detected within the aforementioned astrometric error: a galaxy with coordinates (RA,Dec.)~=~(133.30148,16.43875), companion to the AGN SDSS~J0852.70+16261 in Obs.\#0206340101. Its total 0.3--10~keV count rate is (9$\pm$3)$\times$10$^{-3}$~s$^{-1}$. It belongs to the galaxy pair in our sample corresponding to the smallest separation (6.4"), comparable to the angular resolution of the X-ray telescope. This casts doubts on the robustness of the detection that could be affected by uncertainties in the Point Spread Function modelling (Fig.~\ref{fig:EPIC_vs_SDSS}). For all the other non-AGN galaxies, upper limits on an X-ray counterpart vary in the range between 0.011 and 5~s$^{-1}$ depending on the source position in the EPIC field-of-view and the observation exposure time.

\subsection{XMM-Newton spectroscopy}
\label{sect:spectroscopy}

For each of the X-ray counterparts, we performed a formal forward-folding spectral analysis. Our main goal is to estimate the X-ray luminosity and the column density obscuring the primary emission of the AGN. EPIC source spectra were extracted from a circular region of 20" radius around the SDSS optical position. Background spectra were extracted from boxes in detector coordinates of of 50"$\times$75" and 50"$\times$150" once a circle around the AGN had been removed with a radius comprised between 30" and 40" depending on the source count rate to avoid that the background region is contaminated by the source PSF wings. From the background regions circles were also excised around any serendipitous source (including the position of the non-active SDSS AGN). Instrument responses were generated for each individual source using the standard SAS tasks {\tt rmfgen} and {\tt arfgen} taking into account the corresponding time- and spatial-dependent calibration. 

\begin{table}
\caption{List of models used in the spectroscopic analysis of the AGN counterpart. $A$ is the power-law normalization. $G(E)$ is a Gaussian profile in emission. $C(E)$ describes the emission of an optically-thin, collisionally ionized plasma in thermal equilibrium \citep{smith01}. $R(E)$ is a Compton-reflection model after \citet{magdziarz95}.
}
\begin{tabular}{lc}
\hline
ID & Model \\
\hline
1 & $e^{-\sigma N_H} A_1 E^{-\Gamma}$ \\
2 & $A_1 E^{-\Gamma} + e^{-\sigma N_H} A_2 E^{-\Gamma}$\\
3 & $A_1 E^{-\Gamma} + e^{-\sigma N_H} A_2 E^{-\Gamma} + G(E)$ \\
4 & $C(E) + e^{-\sigma N_H} A E^{-\Gamma}$ \\
5 & $C(E) + R(E) + G(E)$ \\
\hline \hline
\end{tabular}
\label{tab_model}
\end{table}

Spectra were rebinned according to the instrumental resolution following the prescription in \citet{kaastra16}. Spectra were analyzed in the 0.3--10~keV energy range using the Cash statistics \citep{cash76,kaastra17}. We have employed a set of phenomenological models as described in Tab.~\ref{tab_model}\footnote{In the analysis we used a wider set of trial models than shown in Tab.~\ref{tab_model}, which include only those yielding the best-fit on at least one of the sources in Tab.~\ref{tab_fit}.}. They are sufficient to describe the spectral shapes observed in the X-ray spectra of the AGN on our sample. While a detailed physical description of the astrophysical nature of each object is beyond the scope of this paper, the proposed parameterization is sufficient to allow us to estimate the quantities we are interested in (X-ray luminosity and obscuring column density), keeping the systematic uncertainties due to the models lower than the statistical errors. As far as the Compton-reflection model is concerned ($R(E)$ in Tab.~\ref{tab_model}), we used {\tt pexrav} \citep{magdziarz95} for consistency with the results published in the literature on sources of similar statistical quality and for simplicity, even if we are well aware of the existence of more recent and sophisticated models to describe the emission from the AGN torus (see, {\it e.g.}, the discussion in \citealt{murphy09}). As we are not interested in this paper in understanding the detailed properties of the reprocessing material (that would be anyway very hard to determine given the quality of the spectroscopic data) we consider this approximation acceptable. In using {\tt pexrav} we have assumed that the spectral index of the primary illuminating continuum is the same as that of the primary continuum, and that the cut-off energy is 100~keV. This value is consistent with the measurement in several nearby AGN observed with NuSTAR \citep{tortosa18, middei19}. the exact value of this parameter does not have an impact on the results discussed in this paper. Likewise, we employed models of optically-thin, collisionally ionized plasma ({\tt apec}; \citealt{smith01}) to model any "soft excess" component above the extrapolation of the hard X-ray power-law (assumed of nuclear origin) not to unduly over-fit the spectroscopic data, even if there is now overwhelming evidence that any "soft excess" in obscured AGN is due to the emission of photoionized gas in the X-ray Extended NLR \citep{kinkhabwala02,guainazzi07,bianchi19}, while the origin of the soft excess in unobscured AGN is still debated \citep{crummy06,fabian09,done12}. None of these assumptions affects substantially the main conclusions of this {\it paper}.

As the Cash goodness-of-fit test does not allow us to estimate the absolute quality of the fit, the best-fit model was chosen by applying an empirical criterion to models differing by one component corresponding to the the 99\% confidence level of the F-test calculated on the basis of the $\chi^2$ difference between the two models, taking the caveats after \citet{protassov02} into account. We applied the $\chi^2$ on spectra rebinned to a minimum number of 10 counts per background-subtracted spectral channel to minimize biases due to the non-Gaussian nature of the count distribution.

The quality of the X-ray spectroscopic data is insufficient to constrain the primary photon index when: a) the hard X-ray spectrum is dominated by a Compton-reflection component (Model \#5 in Tab.~\ref{tab_model}); b) in a few sources where the primary continuum is heavily obscured and therefore the
number of counts above the photoelectric cut-off energy is too low. In these cases we have assumed a photon index fixed to the value $\Gamma$~=~1.7, the median value observed in a sample of nearby AGN with good quality X-ray spectroscopy data with XMM-Newton \citep{bianchi08}.
In two objects (SDSS~J002920.36-00102 and SDSS~J011254.92+00) the AGN photon index can be measured, but
is marginally inconsistent with this value because it is too hard. We attribute this discrepancy to poor signal-to-noise (in the latter case in presence of
a steeper soft X-ray component). We estimate that the corresponding systematic error on the AGN X-ray luminosity is $\le$20\% if $\Gamma$~=~1.7 is
the true value. In SDSS~J120443.32+103 a flat
spectral index $\Gamma \simeq 1.2$ is measured in a good signal-to-noise spectrum. In this case, we interpret this results as due to a too simple parameterization of the X-ray absorber as a single layer. While more complex X-ray absorption structures are not required by the fit, it is possible that deep X-ray observations at higher energies may unveil an additional, higher column density component. If this is the case, our measurement of the column density in this object, $\simeq$6$\times$10$^{22}$~cm$^{-2}$, is likely to represent a lower
limit to the true column density covering the AGN.

The best-fit values on 2-10~keV observed luminosity, photon index, and column density absorbing the primary continuum are shown in Tab.~\ref{tab_fit}.
\begin{table}
\caption{Best-fit parameters for the X-ray counterparts of the optically classified AGN. $N_H$ is the column density covering the primary continuum; $L_X$ is the 2--10~keV luminosity of the primary continuum corrected for absorption; $\Gamma$ is the photon index of the primary power-law continuum; $\log([OIII])$ is the [OIII]$\lambda5007$ luminosity; $BD$ is the Balmer Decrement; $\log(R)$ is the logarithm of the ratio between the observed 2--10~keV and BD-corrected [OIII] luminosity (cf. Sect.~\ref{sect:optical_spectroscopy}). N/A indicates that the [OIII] measurement is not available. The Model IDs follow Tab.~\ref{tab_model}.}
\begin{tabular}{lcccccccc}
\hline
Source & $\log(L_X)$ & $\log(N_H)$ & $\Gamma$ & $\log(L_{[OIII]})$ & $\log(BD)$ & $\log(R)$ & Model & C/$\nu$ \\
 & & & & & & & & \\
\hline
SDSSJ002920.36-00102 & $42.29\pm^{ 0.38}_{ 0.17}$ & $22.91\pm^{ 0.88}_{ 1.16}$ & $ 0.57\pm^{ 1.07}_{ 1.34}$ & 39.67 & -0.7 &         2.6 & 3  &     103.9/87 \\
SDSSJ010951.34+00024 & $43.55\pm^{ 0.33}_{ 0.50}$ &                $\ge 24.20$ &               1.7$^{\dag}$ & 41.77 &  0.7 &        -0.1 & 5  &      11.5/15 \\
SDSSJ011254.92+000 & $43.49\pm^{ 0.13}_{ 0.28}$ &                $\le 24.25$ & $ 0.28\pm^{ 1.37}_{ 0.87}$ & 39.45 & -2.2 &         4.0 & 2  &    190.2/217 \\
SDSSJ011659.07+00193 & $43.23\pm^{ 0.53}_{ 0.67}$ &                $\ge 24.20$ &               1.7$^{\dag}$ & 41.65 &  1.6 &        -0.3 & 1  &      38.8/25 \\
SDSSJ014402.61-00070 &                $\le 43.35$ &                $\ge 24.20$ &               1.7$^{\dag}$ & 41.56 &  1.4 &   $\le$-0.2 & 5  &      56.1/44 \\
SDSSJ030639.57+00034 & $42.77\pm^{ 0.92}_{ 0.18}$ & $21.55\pm^{ 0.25}_{ 0.74}$ & $ 1.84\pm^{ 0.05}_{ 0.04}$ & 41.34 & -0.2 &         1.4 & 2  &    543.8/494 \\
SDSSJ030655.63-00014 & $44.\pm^{ 0.25}_{ 0.24}$ &                $\ge 24.20$ &               1.7$^{\dag}$ & 42.24 &  1.4 &         0.2 & 5  &      96.7/97 \\
SDSSJ0852.70+16261 & $42.58\pm^{ 0.10}_{ 0.08}$ & $23.00\pm^{ 0.18}_{ 0.20}$ & $ 1.48\pm^{ 0.64}_{ 0.55}$ & 41.05 &  0.4 &         1.5 & 2  &    284.0/277 \\
SDSSJ101858.47+37180 & $43.09\pm^{ 0.23}_{ 0.24}$ &                $\ge 24.20$ &               1.7$^{\dag}$ & 41.43 &  0.6 &        -0.2 & 1  &      95.6/90 \\
SDSSJ104856.96+59282 & $42.85\pm^{ 0.07}_{ 0.07}$ & $21.11\pm^{ 0.19}_{ 0.}$ & $ 1.61\pm^{ 0.18}_{ 0.17}$ & 40.91 &  0.1 &         1.9 & 1  &    263.5/247 \\
SDSSJ115852.20+42432 & $40.66\pm^{ 0.14}_{ 0.14}$ &                $\ge 24.20$ &               1.7$^{\dag}$ & 39.24 &  1.2 &        -0.4 & 4  &    439.4/149 \\
SDSSJ120443.32+103 & $42.61\pm^{ 0.01}_{ 0.02}$ & $22.76\pm^{ 0.04}_{ 0.05}$ & $ 1.23\pm^{ 0.09}_{ 0.11}$ & 41.57 &  0.4 &         1.0 & 3  &    653.4/523 \\
SDSSJ121044.28+38201 & $42.55\pm^{ 0.45}_{ 0.37}$ & $23.17\pm^{ 0.13}_{ 0.13}$ & $ 1.48\pm^{ 0.67}_{ 0.72}$ & 39.80 & -1.0 &         2.7 & 2  &    277.5/304 \\
SDSSJ124210.61+370 & $43.20\pm^{ 0.02}_{ 0.02}$ & $22.06\pm^{ 0.08}_{ 0.06}$ & $ 1.79\pm^{ 0.08}_{ 0.06}$ & 41.45 &  0.0 &         1.8 & 3  &    802.6/498 \\
SDSSJ125729.99+28111 & $42.18\pm^{ 0.10}_{ 0.}$ & $22.48\pm^{ 0.14}_{ 0.13}$ & $ 1.48\pm^{ 1.26}_{ 0.84}$ &   N/A &  N/A &         N/A & 2  &    197.6/185 \\
SDSSJ134736.41+17340 & $41.73\pm^{ 0.24}_{ 0.34}$ &                $\ge 24.20$ &               1.7$^{\dag}$ & 41.44 &  0.5 &        -1.6 & 1  &    195.6/167 \\
SDSSJ145840.74+38273 & $42.21\pm^{ 0.36}_{ 0.52}$ &                $\ge 24.20$ &               1.7$^{\dag}$ &   N/A &  N/A &         N/A & 5  &    214.6/199 \\
SDSSJ170601.87+60173 & $42.27\pm^{ 0.69}_{ 0.71}$ &                $\le 21.82$ & $ 3.53\pm^{ 1.99}_{ 1.09}$ & 41.08 &  0.2 &         1.2 & 1  &      16.9/22 \\
SDSSJ171715.74+64154 & $41.97\pm^{ 0.65}_{ 1.40}$ &                $\ge 24.20$ &               1.7$^{\dag}$ & 41.90 &  1.8 &        -1.8 & 1  &      93.0/64 \\
\hline \hline
\end{tabular}
\label{tab_fit}
\end{table}
We have complemented the results of the X-ray spectral analysis through a study of the ratio between the observed 2--10~keV and [OIII]$\lambda5007$ luminosities. This  is known to be a sensitive diagnostic of X-ray obscuration (\citealt{bassani99}; cf. Sect.~\ref{sect:optical_spectroscopy} for the derivation of the [OIII] luminosity). 
The selection criteria of our sample, extracted from optically-classified AGN/no-AGN pairs (cf. \S~\ref{sect:sample}), ensure that we can use it as a proxy for X-ray obscuration.
The L$_{\rm x}$/L$_{\rm [OIII]}$ ratio has been investigated using different samples of type~1 and type~2 AGN \citep{heckmanetal05, mulchaeyetal94, bassani99, lamastra09, vignalietal10}. 
In particular, in a sample of Compton-thick AGN \cite{marinucci12} measured log~(L$_{\rm x}$/L$_{\rm [OIII]}$)=-0.76  (0.1 dex dispersion), while for Compton-thin AGN \cite{lamastra09} found  log~(L$_{\rm x}$/L$_{\rm [OIII]}$)=1.09 (0.63 dex dispersion).
We will use these thresholds to shed further light on the source classification based on the X-ray spectral analysis.
We assumed a threshold of log~(L$_{\rm x}$/L$_{\rm [OIII]}$)$\le$0.43 to classify a source as Compton-thick.
Out of the {\it bona fide} AGN of our sample, eight sources have a ratio value indicative of Compton-thick obscuration. For these sources we estimate a 2--10 intrinsic luminosity larger than the observed one by a factor of 70, following \citet{lamastra09, marinucci12}. We have also verified {\it a posteriori} that all sources with a log~(L$_{\rm x}$/L$_{\rm [OIII]}$)
between 0.43 and 1.72 are consistent with being Compton-thin according to the X-ray spectral analysis, except one for which only a $\log(N_H) \le 21.82$ upper limit on the X-ray column density (for a rather steep photon index: $\Gamma$$\ge$2.4) could be determined.

We did not perform any spectral analysis on the only X-ray counterpart of the SDSS non-active member of the pairs due to the low statistical quality of the detection (cf. Sect.~\ref{sect:photometry}). Its count rate corresponds to an observed flux of (6$\pm$2)$\times$10$^{-14}$~erg~s$^{-1}$~cm$^{-2}$, or an observed rest-frame luminosity of (5.3$\pm$1.8)$\times$10$^{41}$~erg~s$^{-1}$, for a typical unobscured ($N_H$~=~5$\times$10$^{20}$~cm$^{-2}$) AGN power-law spectrum with a $\Gamma$=2 photon index.

\subsection{Radio emission}
\label{sect:radio}

\begin{figure}
    \centering
    \includegraphics[width=0.75\textwidth]{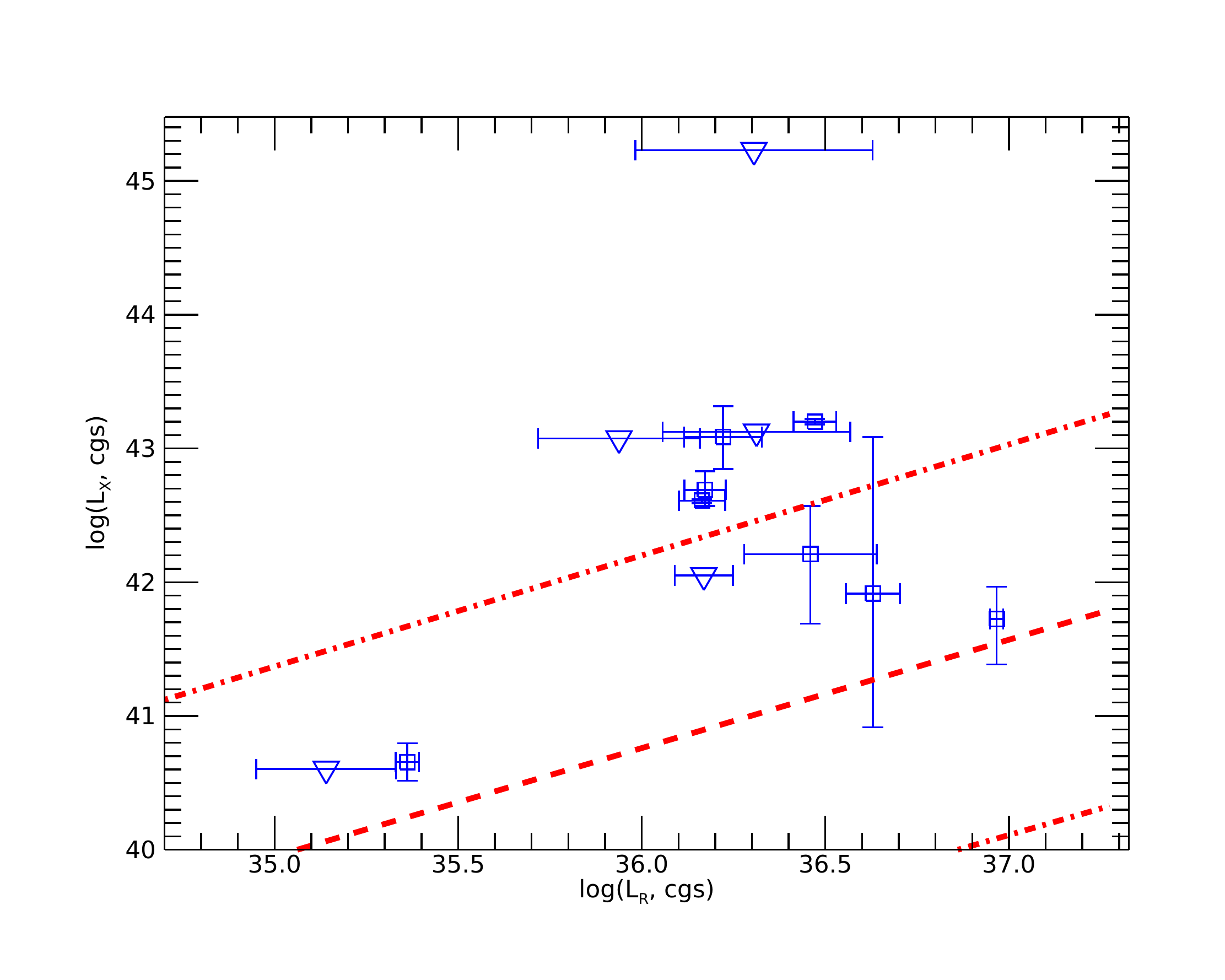}
    \caption{X-ray ($L_x$) versus 1.4~GHz luminosity ($L_R$) for the AGN in the galaxy pair sample discussed in the {\it paper}. {\it Downwards triangles} indicate upper limits on the X-ray luminosity. The {\it red} lines represent the best-fit linear relation ({\it dashed}) and the envelope corresponding to the 1$\sigma$ uncertainty on the best-fit parameters ({\it dot-dashed}) of the relation between the 2-10~keV luminosity and the 20~cm radio luminosity on the sample of local Seyfert galaxies after Panessa et al. (2007).
    }
    \label{fig:radio_vs_x}
\end{figure}

We correlated the position of the AGN in the parent sample of 32 galaxy pairs with the Faint Images of the Radio Sky at Twenty-Centimers (FIRST; \citealt{helfand15}) and the NRAO VLA Sky Survey (NVSS; \citealt{condon98}) 1.4~GHz radio surveys, using a matching radius of $\le$3.5" driven by the X-ray astrometric accuracy. 12 of the X-ray sources exhibit a radio counterpart. A radio source is present at the position of 5 additional AGN with no detected X-ray counterpart. About half of the AGN are X-ray over-luminous with respect to the radio luminosity when compared to the relation derived from a sample of local Seyfert~1 galaxies [$L_X$=(11.60$\pm$0.72)+(0.81$\pm$0.02)$\times$$L_R$] (Fig.~\ref{fig:radio_vs_x}; \citealt{panessa07}).

\section{Discussion}
\label{sect:discussion}

\subsection{The X-ray counterparts of inactive galaxies}

Only 1 out of 32 galaxies optically classified as non-AGN is our sample has a {\it bona fide} X-ray counterpart. It is the companion galaxy of the AGN SDSS~J0852.70+16261. The detection is at the 3$\sigma$ level only (110$\pm$40 total net counts) preventing a detailed spectral analysis. Moreover, it corresponds to the galaxy pair with the smallest separation in our sample (6.4"), comparable to the angular resolution of the X-ray telescope. Assuming it is a {\it bona fide} detection, and a standard unobscured AGN spectrum, its observed rest-frame luminosity ($\simeq$5$\times$10$^{41}$~erg~s$^{-1}$) is consistent with a weak AGN, a heavily obscured Seyfert galaxy, emission from starbursts (although in this case it would correspond to a rather extreme Far Infrared Luminosity of $\sim$10$^{45}$~erg~s$^{-1}$ for the local Universe; \citealt{ranalli03}), or even an extreme Ultra-Luminous X-ray (ULX) source. Its WISE colours (Tab.~\ref{tab:wise}) correspond to an intermediate region where obscured AGN may be located \citep{weston17,ricci17}. They are close to the WISE colours of the pair of X-ray obscured AGN ESO509-IG066 \citep{guainazzi05}.
\begin{table}
\caption{WISE colors for SDSS~J0852.70+16261 and its companion galaxy.
}
\begin{tabular}{lcc}
\hline
WISE name & W1-W2 & W2-W3 \\
J0852.38+162619.7 & 0.355 & 4.064 \\
J0852.87+162615.7 & 0.845 & 3.145 \\
\hline \hline
\end{tabular}
\label{tab:wise}
\end{table}

\begin{figure}
    \centering
   \includegraphics[width=0.5\textwidth, angle=90]{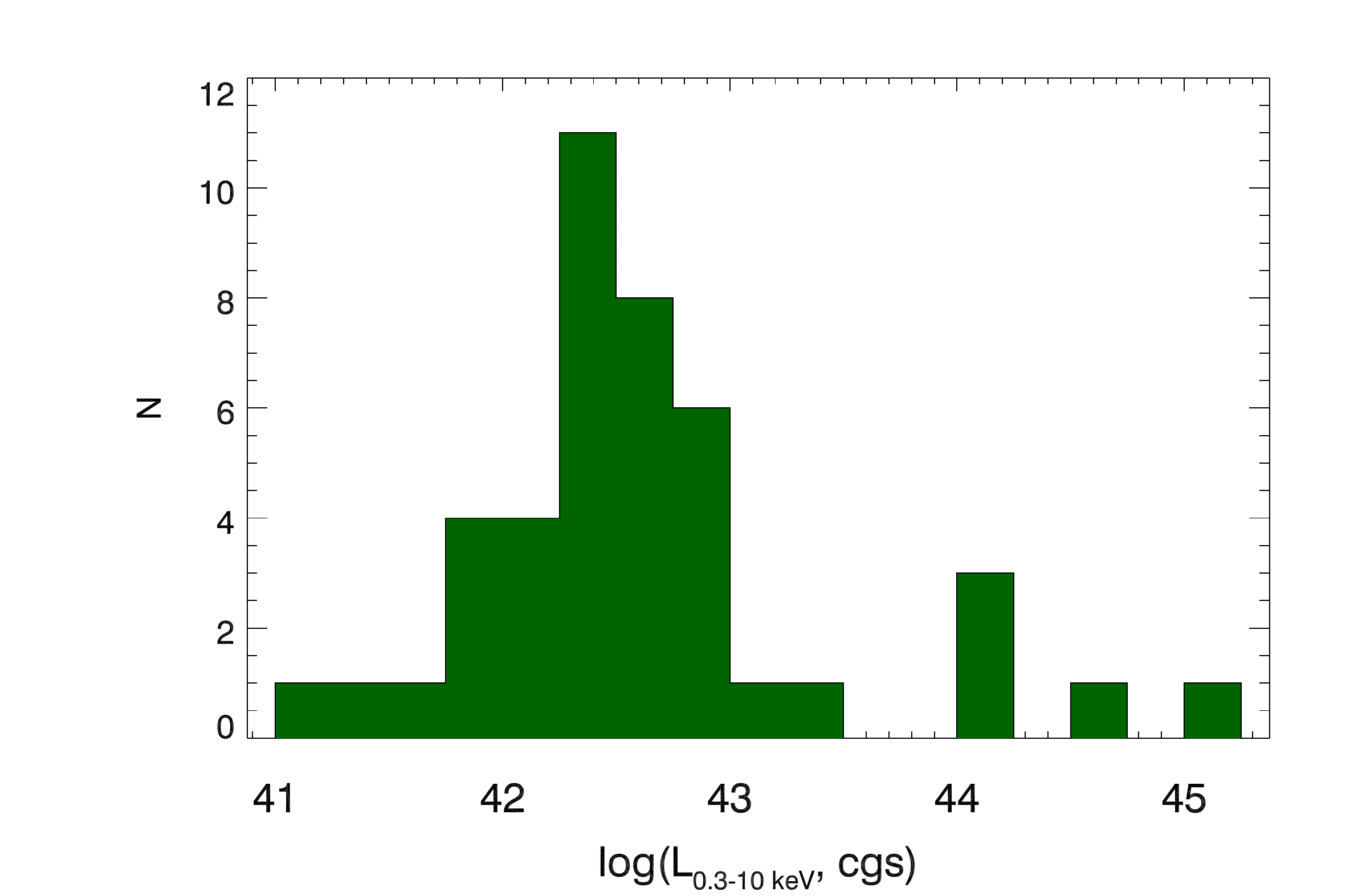}
    \caption{90\% confidence level upper limits on the 0.3--10~keV observed rest frame luminosity at the position of the galaxies optically classified as "non-AGN" is our sample.}
    \label{fig:nagn_ul}
\end{figure}

The nature of the X-ray emission in galaxies without detected X-ray counterparts remains elusive. In Fig.~\ref{fig:nagn_ul}
we show the upper limits on the 0.3--10~keV observed rest frame luminosity at the position of the galaxy optical centroid.
In all cases, the upper limits exceed 10$^{41}$~erg~s$^{-1}$, and 10$^{42}$~erg~s$^{-1}$ for $>$90\% of the sample. Deeper observations would be needed to either detect AGN still confused in the observation noise, or to rule out with certainty the presence of weak of heavily obscured AGN in these optically inactive galaxies. Further observations
with {\it Chandra} would also surely help in ensuring that weak sources are not confused by the comparatively
large and complex PSF of the XMM-Newton telescope.

\subsection{AGN triggering in AGN/no-AGN galaxy pairs}

In order to make a statistical analysis of the obscuration in merging galaxies, we collected from literature the systems hosting AGN/no-AGN pairs and for which a measure of the absorption column density can be derived from X-ray observations. We selected from literature additional 14 pairs AGN-galaxy
in different wavebands, \citep{satyapal17,ricci17,pfeifle19,hou19}. Together with our systems with measured value of N$_{\rm H}$, we collected a total sample of 33 systems (see Tab.~\ref{tab_fit},
and Tab.~\ref{nh_lit}).
We compared this enlarged sample with both a sample of dual AGN (the $\simeq$60 objects extracted from an heterogeneous literature sample by \cite{derosa18}; see also \cite{ricci17}) and with a sample of isolated AGN
(728 non blazar AGN extracted from the 70-month catalogue of the {\it Swift}/BAT; \cite{ricci15}) in the same range of X-ray luminosities (5$\times$10$^{40}$--2$\times$10$^{44}$~erg~s$^{-1}$).

\begin{figure}
    \centering
   \includegraphics[width=0.75\textwidth]{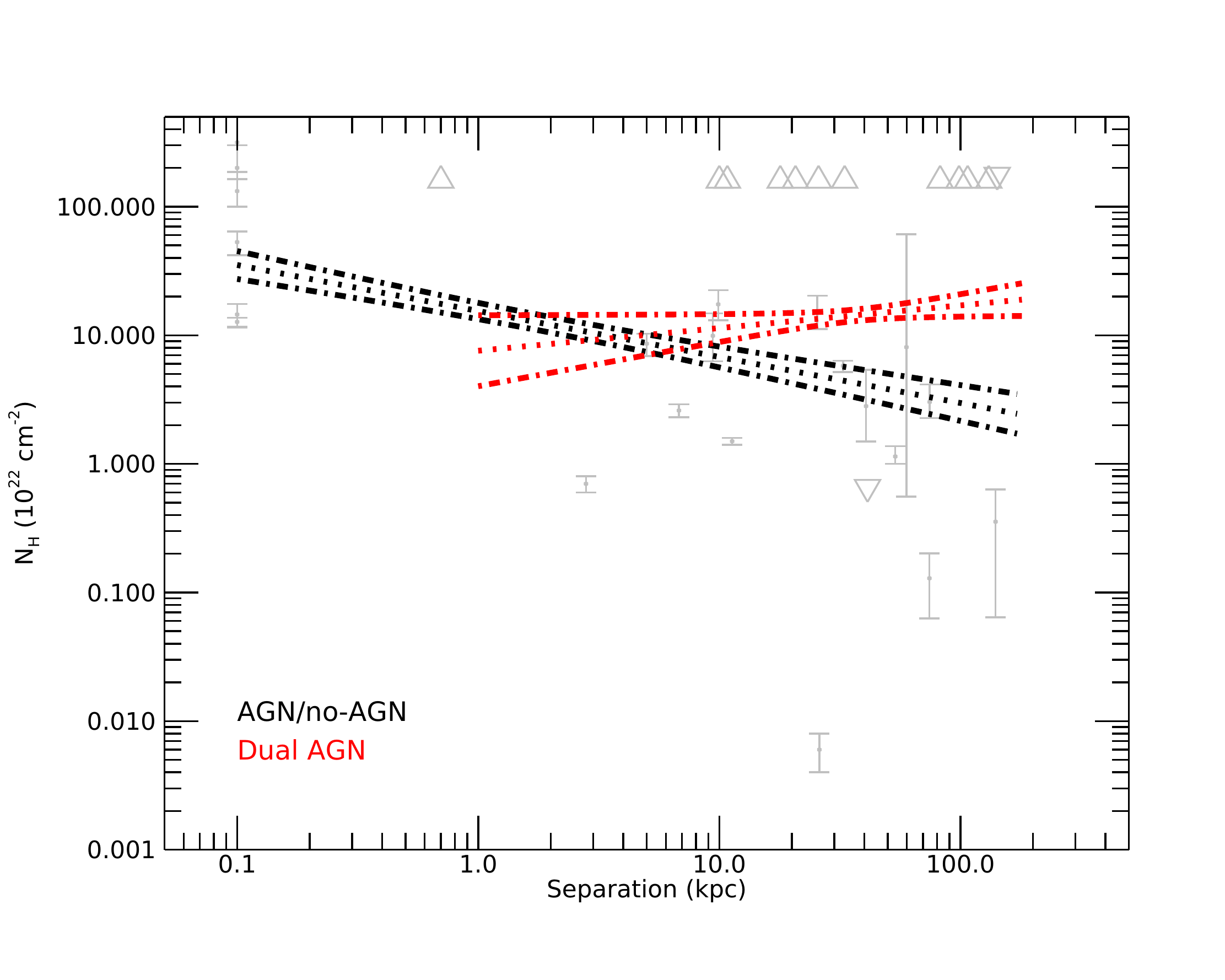}
    \caption{X-ray nuclear column density versus galaxy separation for the AGN/no-AGN sample. The {\it light silver} data points 
    represent the measurements presented in this paper augmented of the literature systems listed in Tab.~\ref{nh_lit}. Upwards (downwards) triangles represent lower (upper) limits. The {\it black lines} represent the best-fit ({\it dotted} and the 
    envelope corresponding to the 1$\sigma$ statistical uncertainties (details of the fit procedure in text). The {\it red lines}
    represent the same fit for a sample of {\it bona fide} dual AGN (data not shown).}
    \label{fig:nh_vs_sep}
\end{figure}

In Fig.~\ref{fig:nh_vs_sep} we show the evolution of the X-ray column density as a function of galaxy separation. The data points
represent the AGN/no-AGN sample. Given the large number of censored data, we fit the data with a Monte-Carlo method, whereby 10,000
sets of simulated data sets were drawn from the observed data set by: a) replacing each column density constrained measurement with the sum of the best-fit value and the statistical error multiplied by a random number extracted from a Gaussian distribution of mean zero and unitary standard deviation; b) replacing each column density censored measurement with a value randomly extracted from a
uniform distribution in the [LL,10$^{19}$~cm$^{-2}$] and [UL, 10$^{26}$~cm$^{-2}$] for Lower Limits (LL) and Upper Limits (UL), respectively. The curves in Fig.~\ref{fig:nh_vs_sep} represent the average of the fit on the 10,000 simulated data sets ({\it dotted})
and the 1$\sigma$ envelope ({\it dashed-dotted)} of the best-fit curves distribution. Formally, a strong anti-correlation between X-ray obscuration and
galaxy separation is found, with average $N_H$ decreasing from a few 10$^{23}$~cm$^{-2}$ to $\sim$10$^{22}$~cm$^{-2}$ from the sub-kpc to the $\sim$100~kpc scale. However, a few caveats shall be borne in mind. The correlation is driven by a few heavily obscured sources at sub-kpc scale. If they are removed from the fit, the anti-correlation between X-ray obscuration and galaxy separation in the AGN/no-AGN systems becomes statistically not significant. Furthermore, at separations larger than 10~kpc a sizeable number of Compton-thick objects are found. However, the formal anti-correlation is only marginally affected by the inclusion/removal of the Compton-thick sources (due to their homogeneous distribution along the separation axis). 

Out of the 19 sources in our sample for which measurements of the Balmer Decrement (BD) and of the X-ray obscuration are available,
7 out of 9 objects with $BD \ge$~2, and all the 7 objects with $BD \ge$~3 are X-ray Compton-thick. For lower values of the Balmer Decrement no correlation with the
X-ray column density is found. This may indicate that the highest obscuration in the NLR of these systems is
associated with the most compact structures surrounding the active nuclei, possibly the torus. No constraints on the location
of the Compton-thin obscurer can be derived from the data our our sample. It is well possible that Compton-thick and -thin
absorbers in the AGN our our sample probe entirely different systems, thus complicating the interpretation of the correlation
shown in Fig.~\ref{fig:nh_vs_sep}.

With this caveat in mind, the red curve in Fig.~\ref{fig:nh_vs_sep} represents the results of the same fit procedure applied to
the dual AGN sample. In the common separation range ($\ge$10~kpc), dual AGN exhibit an X-ray obscuration higher by above one
order of magnitude than the AGN/no-AGN pairs. The lack of dual AGN corresponding to sub-kpc galaxy separation prevents us from
extending this comparison to the most compact systems.
The total fraction of "obscured" AGN ($N_H \ge$10$^{22}$~cm$^{-2}$) in the AGN/no-AGN sample is
comprised between 79\% and 84\% (cf. Tab.~\ref{tab_fit}). This is significantly larger than observed in
isolated AGN of comparable nuclear luminosity ($\simeq$45\%; \citealt{ricci15}). It is, however, comparable to the fraction
of obscured AGN in a dual AGN sample that raises slowly from $\simeq$70\% at galaxy pair
separations larger than 60~kpc to almost 90\%
for separations lower than 20~kpc.

\begin{figure}
    \centering
   \includegraphics[width=0.75\textwidth]{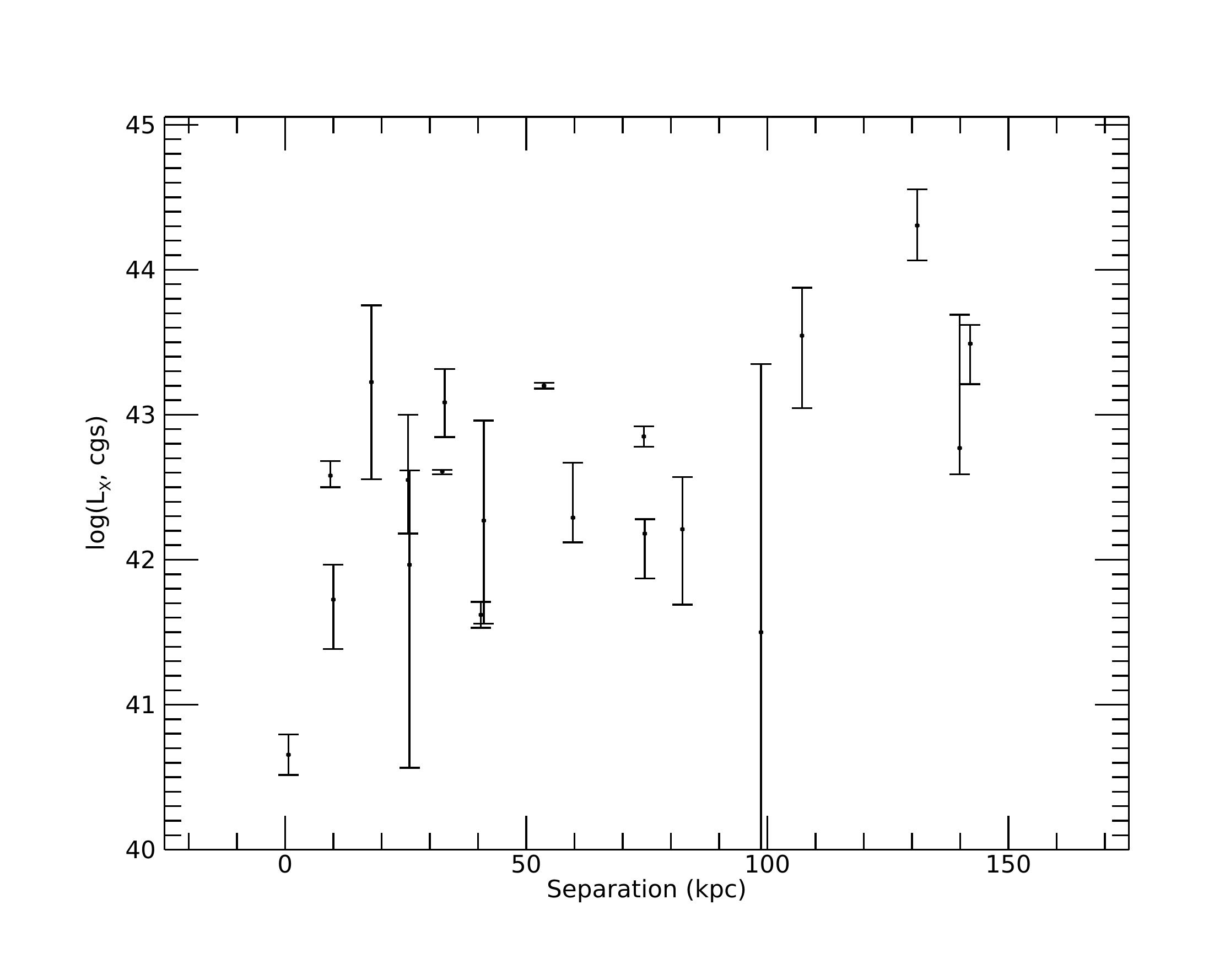}
    \caption{X-ray nuclear luminosity (in logarithm of cgs units) as a function of galaxy pair separation in the AGN/no-AGN sample presented in this {\it paper}.}
    \label{fig:lx_vs_sep}
\end{figure}

In Fig.~\ref{fig:lx_vs_sep} we show the nuclear X-ray luminosity as a function of galaxy separation for the {\it bona fine} AGN
presented in this {\it paper}. A linear fit with a function: $\log(L_X) = A + B \times s$ (where $s$ is the galaxy pair separation
in kpc) yields a moderate trend for more compact systems to be less luminous: $A=42.05 \pm 0.03$, and
$B=0.0178 \pm 0.0008$~kpc$^{-1}$, corresponding to a decrease of 1~dex in luminosity per $\sim$60~kpc. However, it should be
borne in mind that our sample is neither complete nor unbiased in any sense. It is therefore hard to assess the significance
of such a weak trend on the unknown parent population.

Both cosmological \citep{steinborn16,volonteri16} and idealised hydrodynamic simulations \citep{blecha16,capelo17} of merging galaxies hosting BH have been extensively investigated in order to evaluate the main parameters triggering the AGN activity in interacting galaxies. In particular, numerical simulations  show an increase of $N_{\rm H}$ as the distance decreases, due to merger dynamics. The simulation predict a median value of $N_{\rm H}$ of $\sim$ 3$\times$10$^{23}$\cm2, that is in good agreement with the value measured in sample of dual AGN \citep{derosa18}.
It worth noticing that the spatial resolution of the simulation only allow to probe absorption on relatively large scales (above $\sim$ 50~pc), this means that absorption from the AGN torus, on typical pc-scale, is not considered and then the \nh\ value from simulations represents a lower limit.

Simulations of the dynamical evolution of isolated mergers indicate that the evolution depends in principle on several parameters of these complex systems: galaxy
morphology, bulge-to-disk stellar mass ratio, orbital configuration, galaxy mass ratio, bulge-to-disk mass ratio, galaxy mass ratio, black hole mass ratio, and the black hole feedback efficiency \citep{vanwassenhove12,blecha16,blecha18,capelo17}.
All simulations in general agree on the dual AGN phase to
be short, primarily concentrated when the two black holes are separated by less then $\sim$10~kpc.
For typical AGN luminosities $\ge$10$^{42}$~erg~s$^{-1}$, pairs of merging galaxies are supposed
to spend just a few Myr in a "dual AGN" phase \citep{capelo15}. At black hole separations $\ge$10~kpc,
the fraction of the dual active phase versus the total active phase ({\it i.e.}, the phase
when at least one of the AGN is actively accreting) goes from a few percent to about 20 percent. As our sample was pre-selected as galaxy pairs constituted by an
active and an inactive galaxy (according to an optical spectroscopic classification), it does not
allow a direct comparison with the predictions of merging simulations. However, when this comparison
is possible ({\it e.g}, against the all-sky {\it Swift}-BAT survey; \citealt{koss12}) consistency
between the fraction of dual AGN in a galaxy sample and the simulation predictions within a factor of two was found.

\begin{table}
\caption{Separations, $s$, and column densities, $N_H$ for the additional AGN/no-AGN systems extracted from the literature.}
\begin{tabular}{lcc}
\hline
Source & $s$ & $\log(N_H)$ \\
 & (kpc) & \\
\hline
IRAS05189-2524       &  0.1 & $23.10\pm^{ 0.03}_{ 0.04}$ \\
Mrk2               &  0.1 & $23.16\pm^{ 0.08}_{ 0.10}$ \\
NGC34                &  0.1 & $23.72\pm^{ 0.08}_{ 0.10}$ \\
UGC05101             &  0.1 & $24.12\pm^{ 0.09}_{ 0.12}$ \\
NGC7130              &  0.1 & $24.30\pm^{ 0.18}_{ 0.30}$ \\
IRAS120-5453       &  0.1 & $24.50\pm^{ 0.24}_{ 0.23}$ \\
J1036+0221           &  2.8 & $21.85\pm^{ 0.06}_{ 0.07}$ \\
J2356-1016           &  5.0 & $22.93\pm^{ 0.08}_{ 0.10}$ \\
J1147+0945           &  6.8 & $22.41\pm^{ 0.05}_{ 0.05}$ \\
J0859+10           &  9.9 & $23.24\pm^{ 0.11}_{ 0.12}$ \\
NGC4922N             & 10.8 &                $\ge 24.20$ \\
CG468-002W           & 11.3 & $22.18\pm^{ 0.03}_{ 0.03}$ \\
NGC7674              & 20.7 &                $\ge 24.20$ \\
NGC7469              & 26.0 & $19.78\pm^{ 0.12}_{ 0.18}$ \\
\hline \hline
\end{tabular}
\label{nh_lit}
\end{table}


\section{Conclusions}
\label{sect:conclusions}

In this paper, we present an XMM-Newton study of a sample
of galaxy pairs, identified through SDSS spectroscopy as being constituted by an AGN-hosting and an inactive galaxy.
The original main goal of the project was to identify via X-rays active nuclei in the "inactive" member of the pair that could have been missed due to heavy obscuration along the line of sight to the nucleus. For the galaxies hosting an AGN, we discuss in this paper the X-ray photometric and spectroscopic properties, and put them in context of the prediction of state-of-the-art hydrodynamical simulations of isolated mergers on the condition for the triggering of "dual" AGN.

Out of the the parent sample of over 1500 galaxy pairs with an SDSS "AGN--no-AGN" classification, 32 {\it bona fide} pairs are in the field-of-view of the EPIC cameras (as of March 2018). The main results of our study can be summarized as follows:

\begin{itemize}
\item for only one of the "non-AGN" galaxy in the pair an X-ray counterpart is found (at the 3$\sigma$ level). Assuming a standard AGN-like X-ray spectrum, its rest-frame luminosity ($\simeq$5$\times$10$^{41}$~erg~s$^{-1}$) is consistent with a weak AGN, a heavily-obscured Seyfert galaxy (consistently with its WISE colors), emission from starbursts, or even an extreme ULX source

\item for 90\% of the "non-AGN galaxies" in the pair for which no X-ray counterpart is found, the derived upper limits on the 2-10~keV luminosities are $\ge$10$^{42}$~erg~s$^{-1}$, {\it i.e.} still consistent with an active nucleus "hidden" in the data (cf. Fig.~\ref{fig:nagn_ul})

\item for 19 "AGN" galaxies in the pair, a {\it bona fide} X-ray counterpart is found, covering a wide range in absorption corrected X-ray luminosity (5$\times$10$^{40}$--2$\times$10$^{44}$~erg~s$^{-1}$;
cf. Tab.~\ref{tab_fit})

\item our AGN/no-AGN sample shows only a weak dependence of the AGN X-ray luminosity on the separation between the members of the galaxy pair, with more compact systems being marginally {\it less luminous} (cf. Fig.~\ref{fig:lx_vs_sep})

\item the {\it bona fide} AGN are unevenly split between "unobscured" ($N_H$$<$10$^{22}$~cm$^{-2}$; 16--26\%), "Compton-thin ($10^{22}$$\le$$N_H$ $<$1.6$\times10^{24}$~cm$^{-2}$; 32-37\%), and "Compton-thick" ($N_H \ge 1.6 \times 10^{24}$~cm$^{-2}$; 47\%)
obscured (cf. Tab.~\ref{tab_fit})

\item we considered 14 additional systems AGN--non-AGN from literature and compare the total of 33 (14+19) systems with a sample of {\it bona fide} dual AGN \citep{derosa18}. Dual AGN at separations larger than 10~kpc exhibit on the average a larger nuclear obscuration by about one order
of magnitude with respect to the AGN/no-AGN systems (cf. Fig.~\ref{fig:nh_vs_sep})

\item galaxy pairs including {\it at least} an AGN exhibit a fraction of obscured nuclear activity larger than 70\%. This exceeds by a factor 1.5 
the fraction of obscured objects in a sample of isolated AGN of matching luminosities and redshifts \citep{ricci15}
 
\end{itemize}

While this is suggestive that the galactic environment has a key influence on the AGN triggering in merging galaxies, deeper studies with a larger number of objects would be needed in order to perform a more quantitative comparison with the X-ray predictions of hydrodynamic simulations \citep{capelo15,capelo17}. We aim at continuing these studies in the future \citep{derosa19}.

\section*{Acknowledgements} The authors thanks an anonimous referee, whose careful revision of the manuscript significantly improved the paper. We thank G. Calderone for help with QSFit. SB, ADR and EP acknowledge financial support from the Italian Space Agency under grant ASI-INAF 2017-14-H.O.
MPT acknowledges support by the Spanish MCIU through grant PGC2018-098915-B-C21 cofunded
with FEDER funds and from the State Agency for Research of the Spanish MCIU through the ``Center of 
Excellence Severo Ochoa'' award for the Instituto de Astrof\'{\i}sica de Andaluc\'{\i}a (SEV-2017-0709) and through grant PGC2018-098915-B-C21 (MCI/AEI/FEDER, UE).
Funding for SDSS-III has been provided by the Alfred P. Sloan Foundation, the Participating Institutions, the National Science Foundation, and the U.S. Department of Energy Office of Science. The SDSS-III web site is http://www.sdss3.org/.
SDSS-III is managed by the Astrophysical Research Consortium for the Participating Institutions of the SDSS-III Collaboration including the University of Arizona, the Brazilian Participation Group, Brookhaven National Laboratory, Carnegie Mellon University, University of Florida, the French Participation Group, the German Participation Group, Harvard University, the Instituto de Astrofisica de Canarias, the Michigan State/Notre Dame/JINA Participation Group, Johns Hopkins University, Lawrence Berkeley National Laboratory, Max Planck Institute for Astrophysics, Max Planck Institute for Extraterrestrial Physics, New Mexico State University, New York University, Ohio State University, Pennsylvania State University, University of Portsmouth, Princeton University, the Spanish Participation Group, University of Tokyo, University of Utah, Vanderbilt University, University of Virginia, University of Washington, and Yale University. 

\section*{Appendix A: XMM-Newton/EPIC spectra of {\it bona fide} AGN}

In Fig.~\ref{fig:spectra1} and \ref{fig:spectra2} we show the EPIC spectra, best-fit model ({\it upper panels}) and
residuals against the best-fit model in units of data/model ratio for the 19 {\it bona fide} AGN in our sample. each data point correspond to a signal-to-noise ratio larger than three.
\begin{figure*}
    \centering
    \hbox{
   \includegraphics[width=0.225\textwidth, angle=-90]{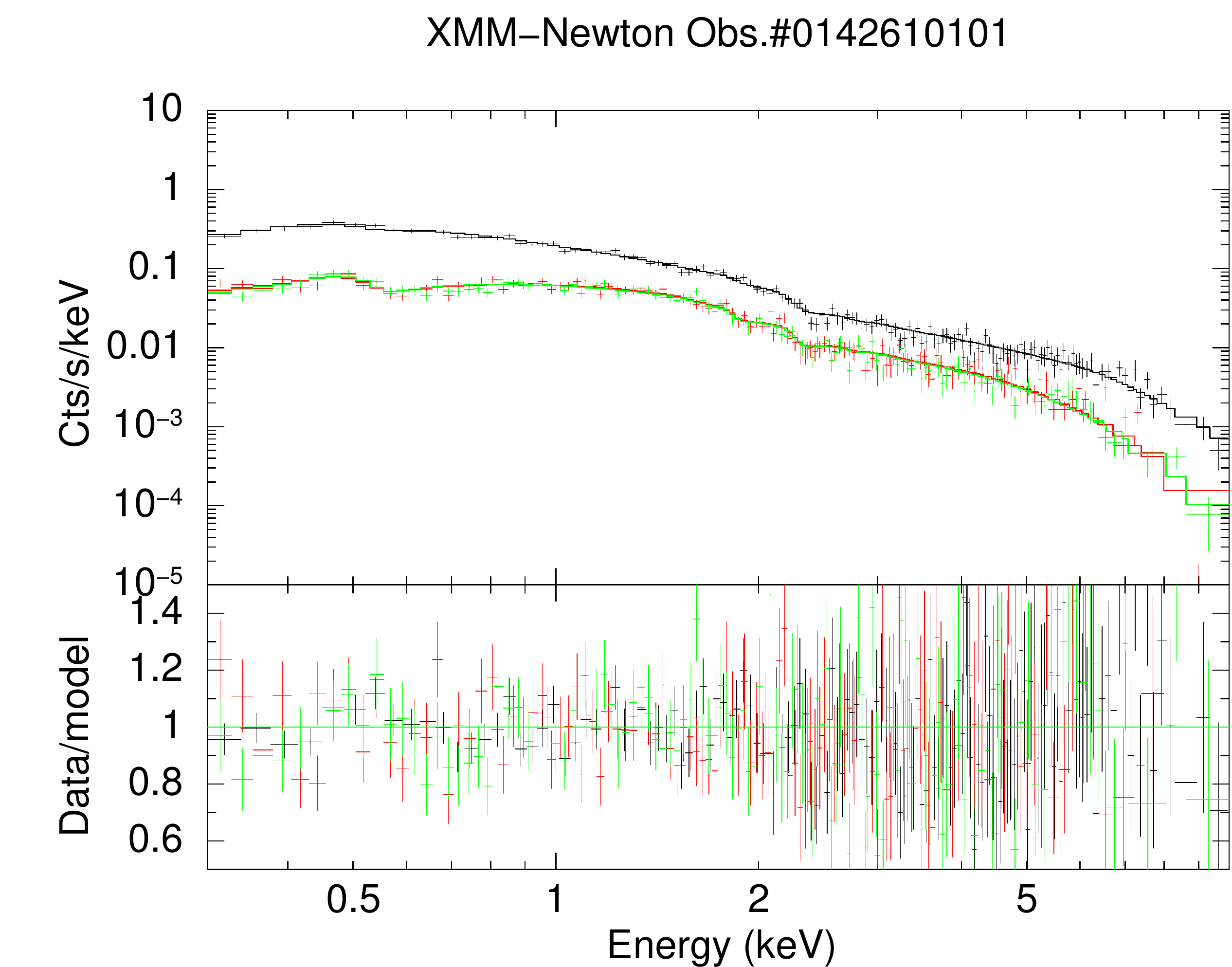}
   \includegraphics[width=0.225\textwidth, angle=-90]{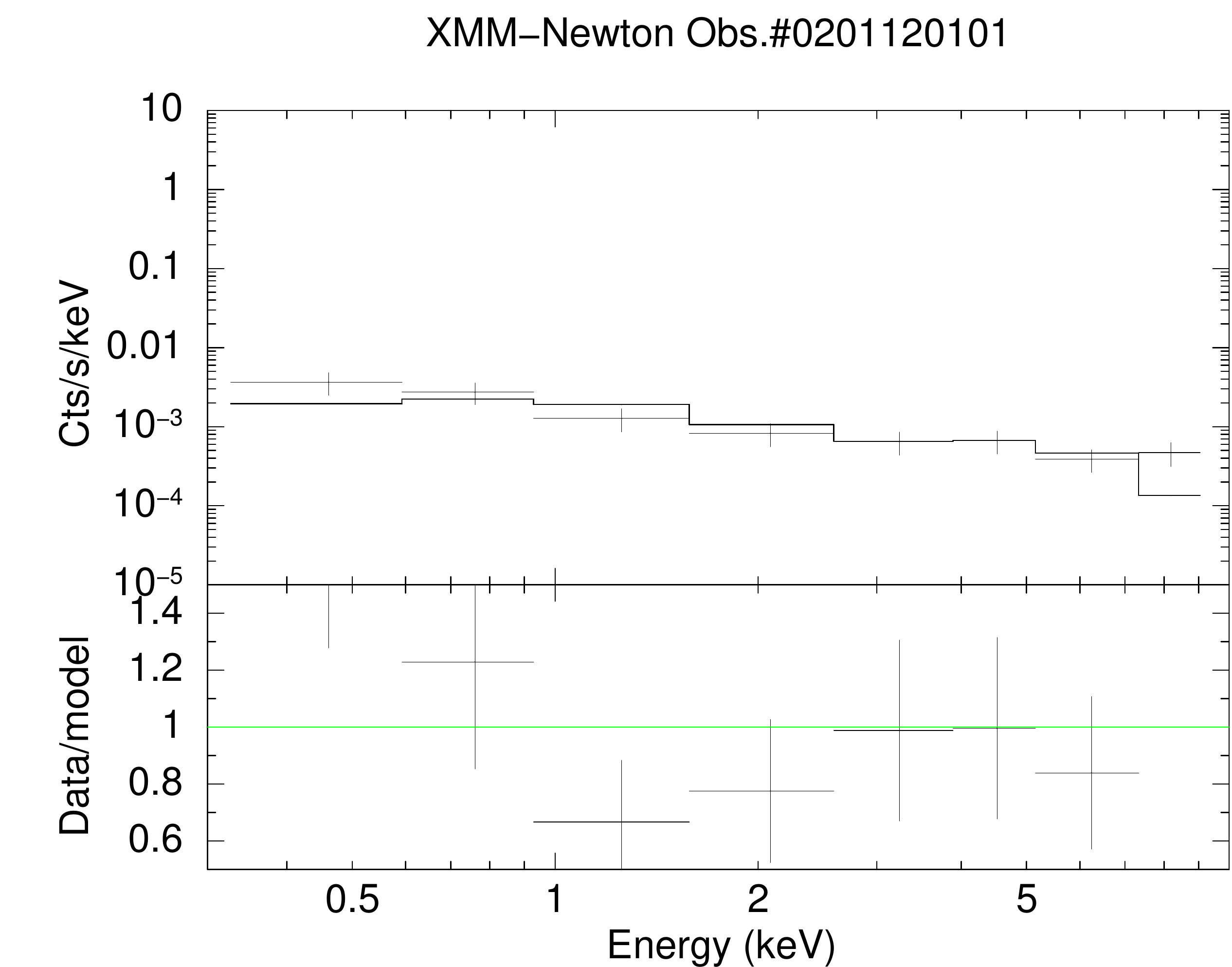}
   \includegraphics[width=0.225\textwidth, angle=-90]{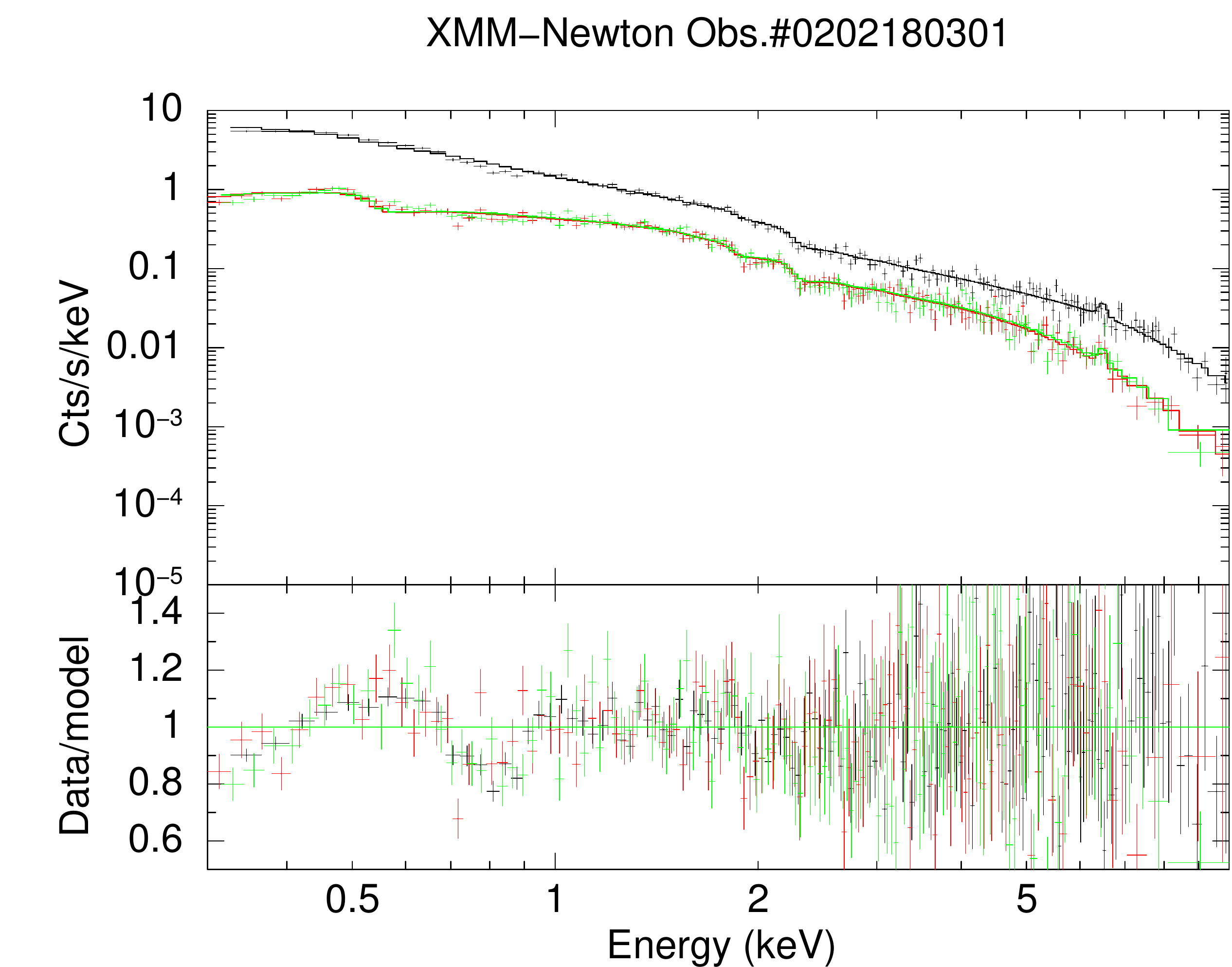}
   }
   \hbox{
   \includegraphics[width=0.225\textwidth, angle=-90]{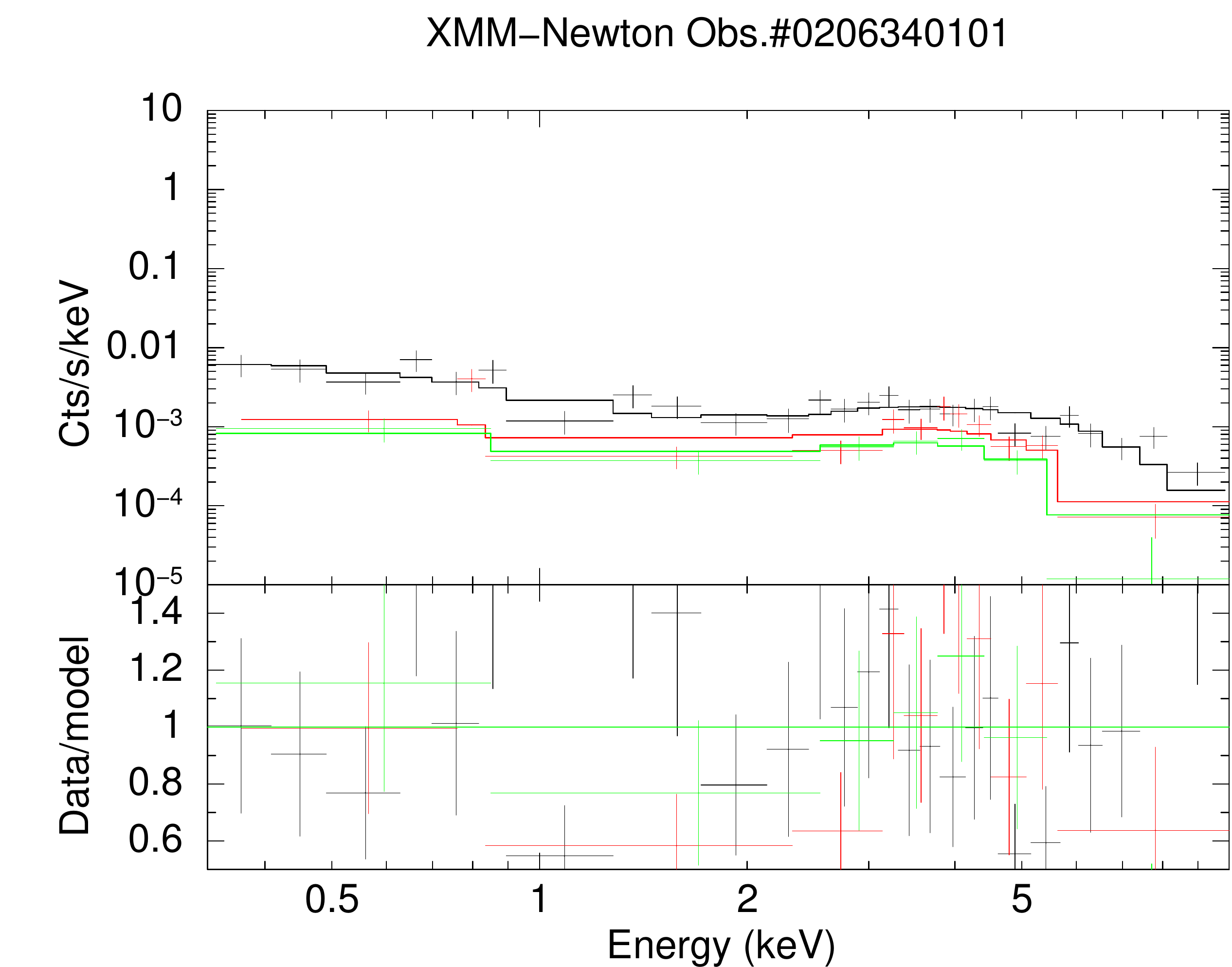}
   \includegraphics[width=0.225\textwidth, angle=-90]{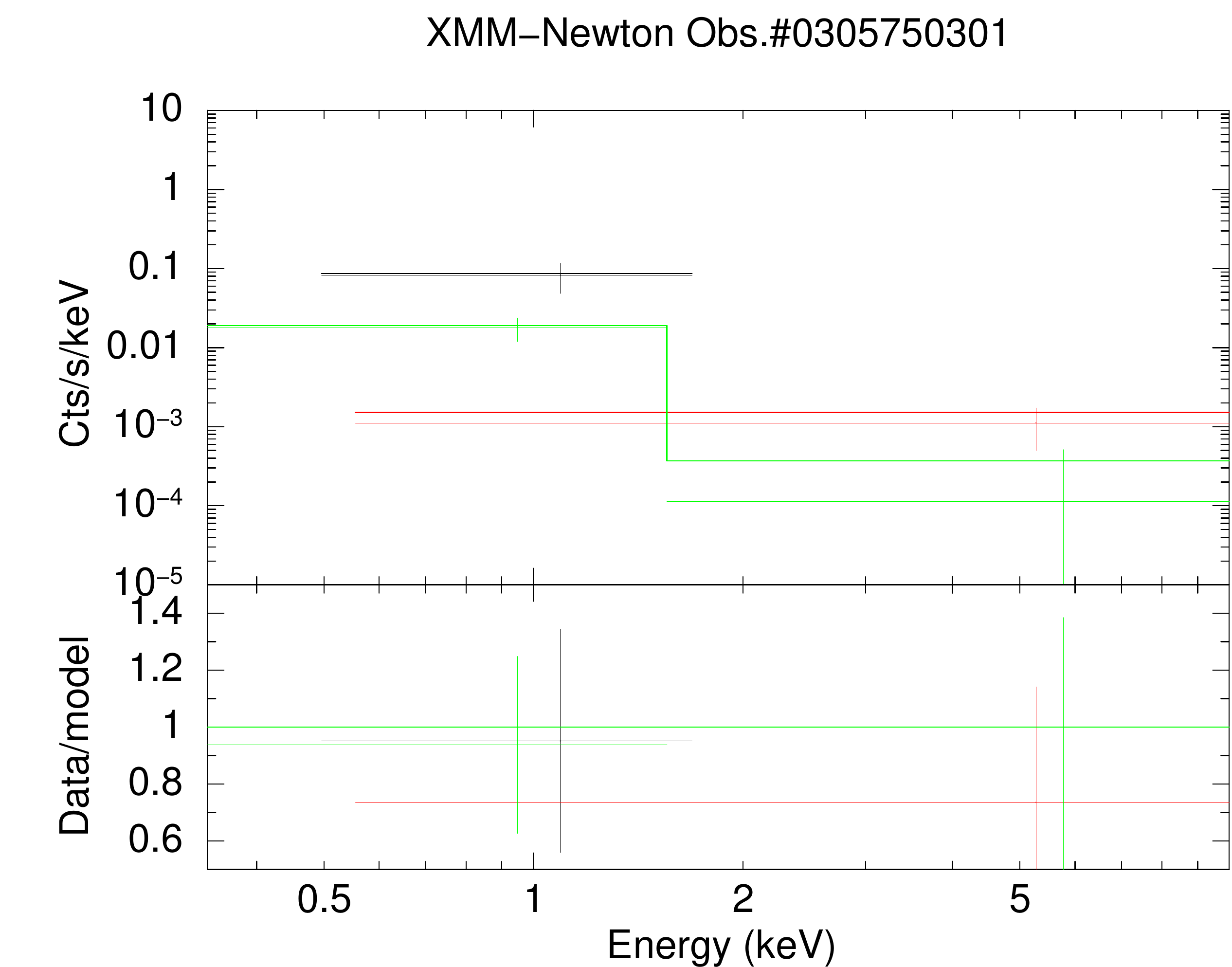}
   \includegraphics[width=0.225\textwidth, angle=-90]{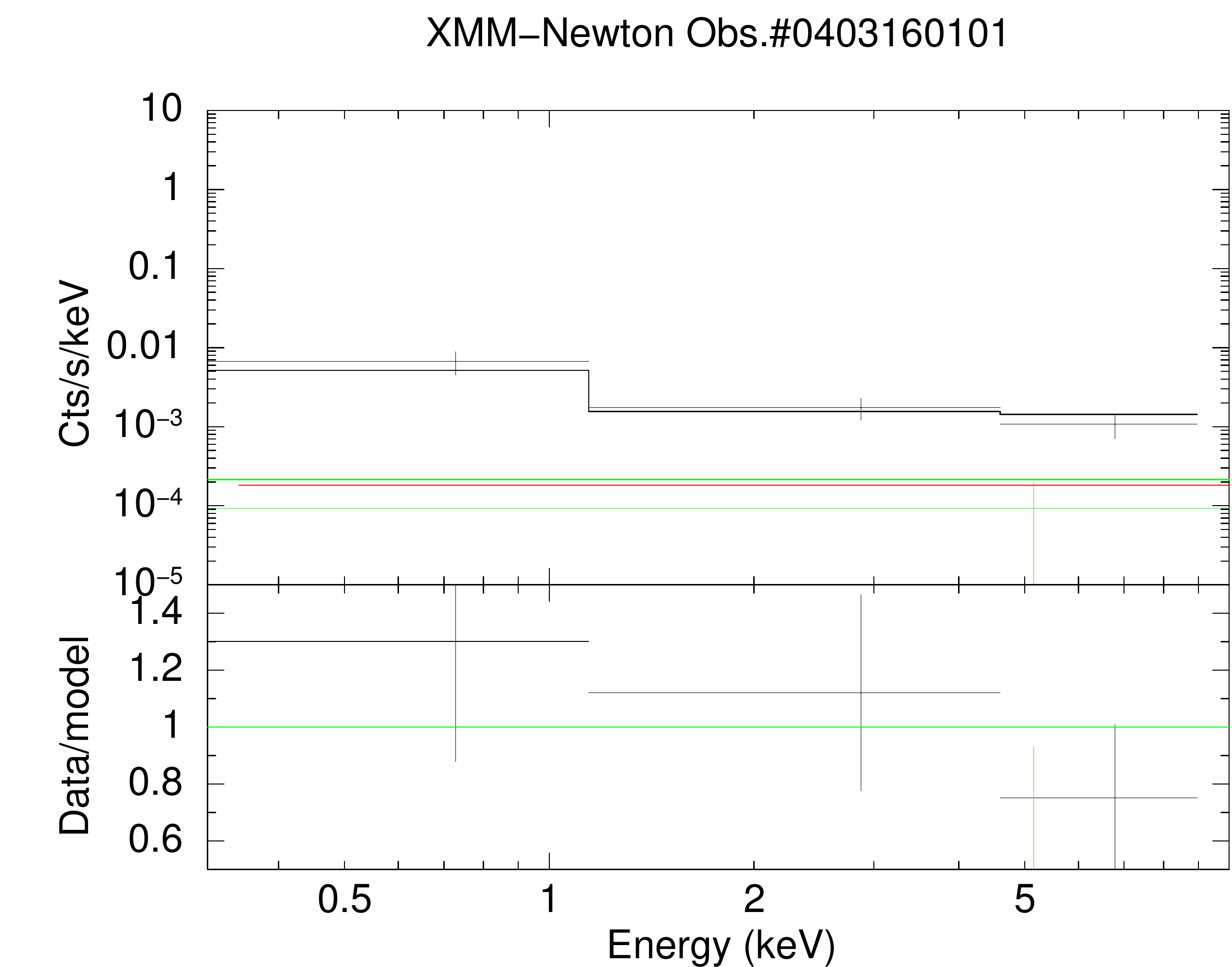}
   }
   \hbox{
   \includegraphics[width=0.225\textwidth, angle=-90]{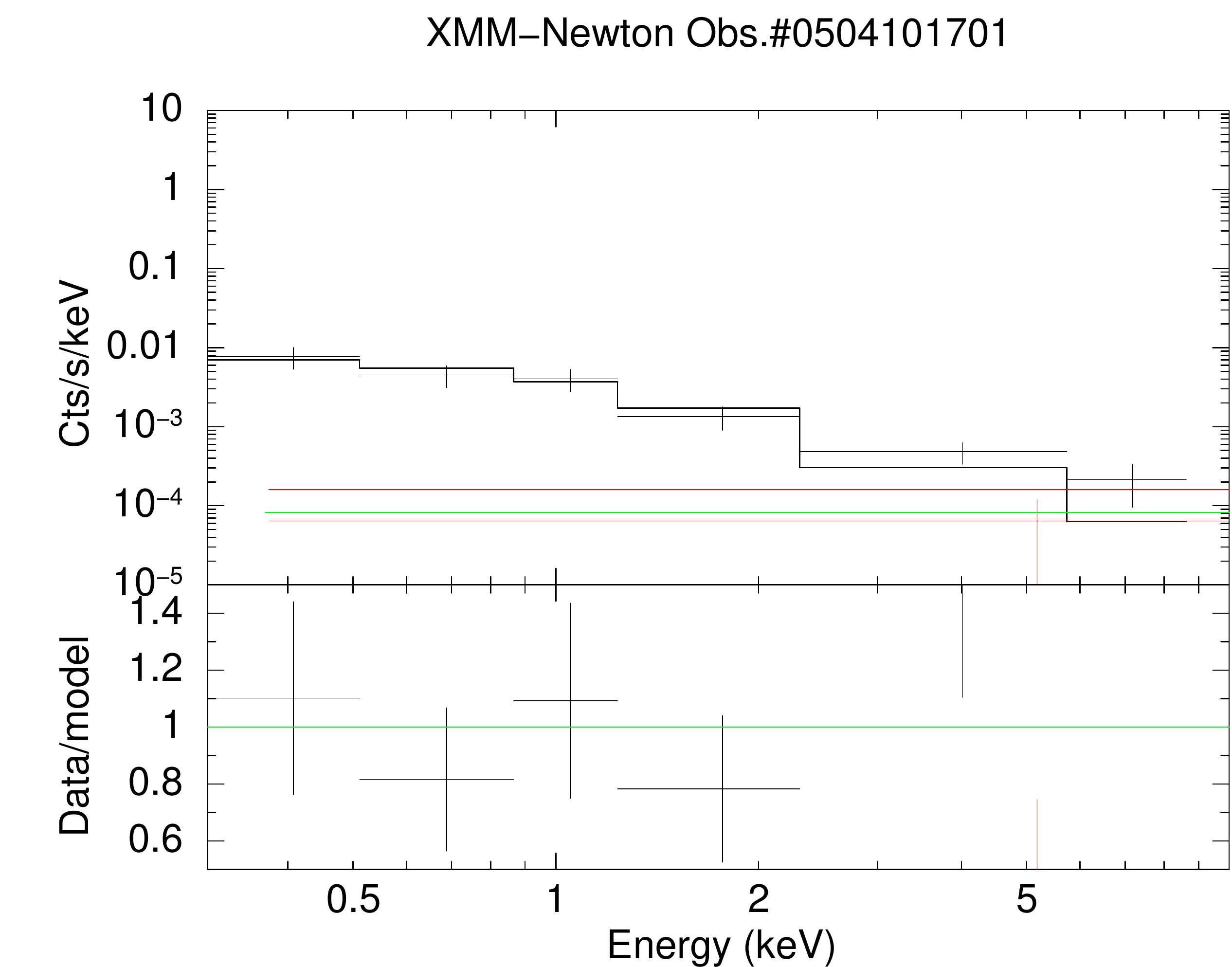}
   \includegraphics[width=0.225\textwidth, angle=-90]{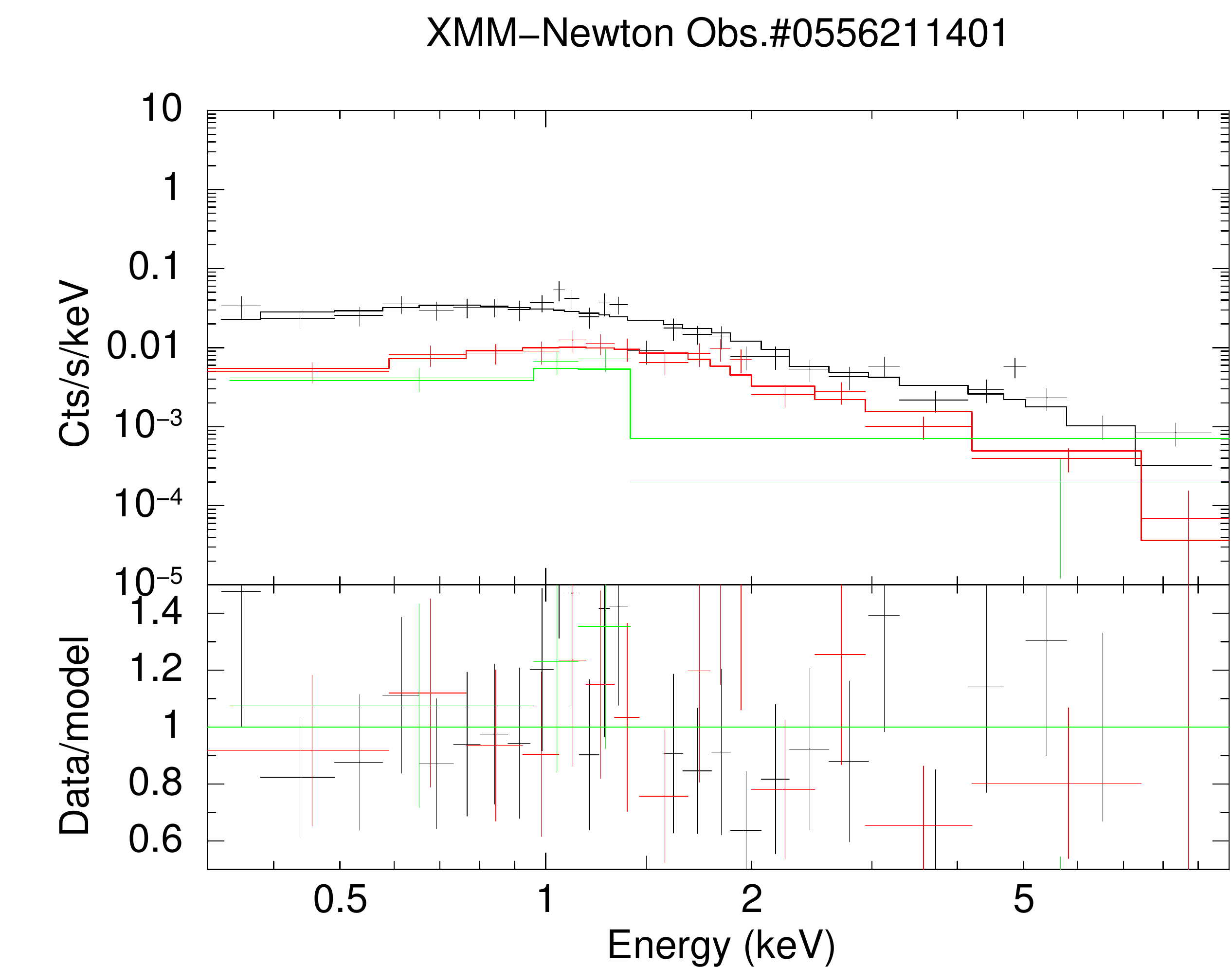}
   \includegraphics[width=0.225\textwidth, angle=-90]{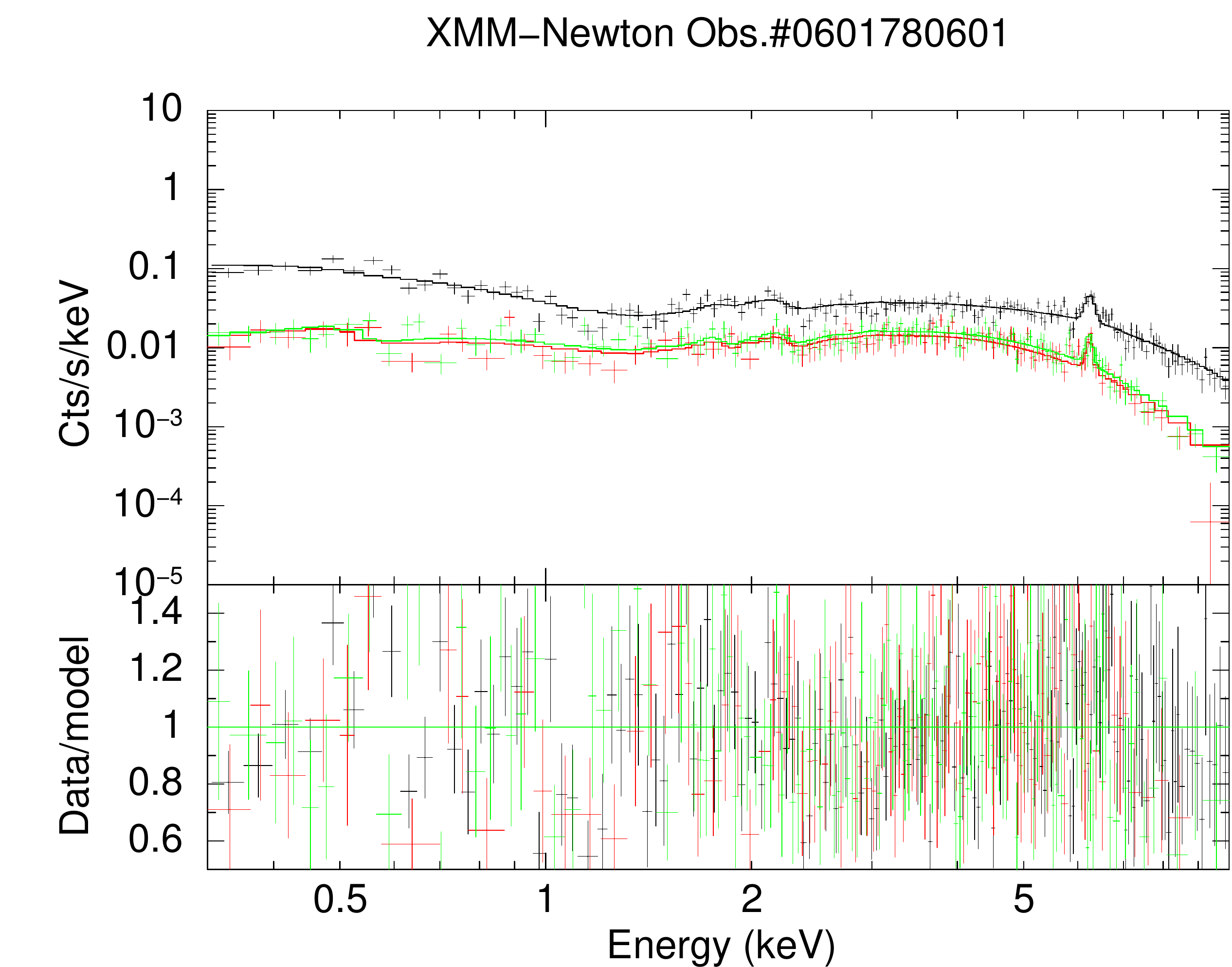}
   }
   \hbox{
   \includegraphics[width=0.225\textwidth, angle=-90]{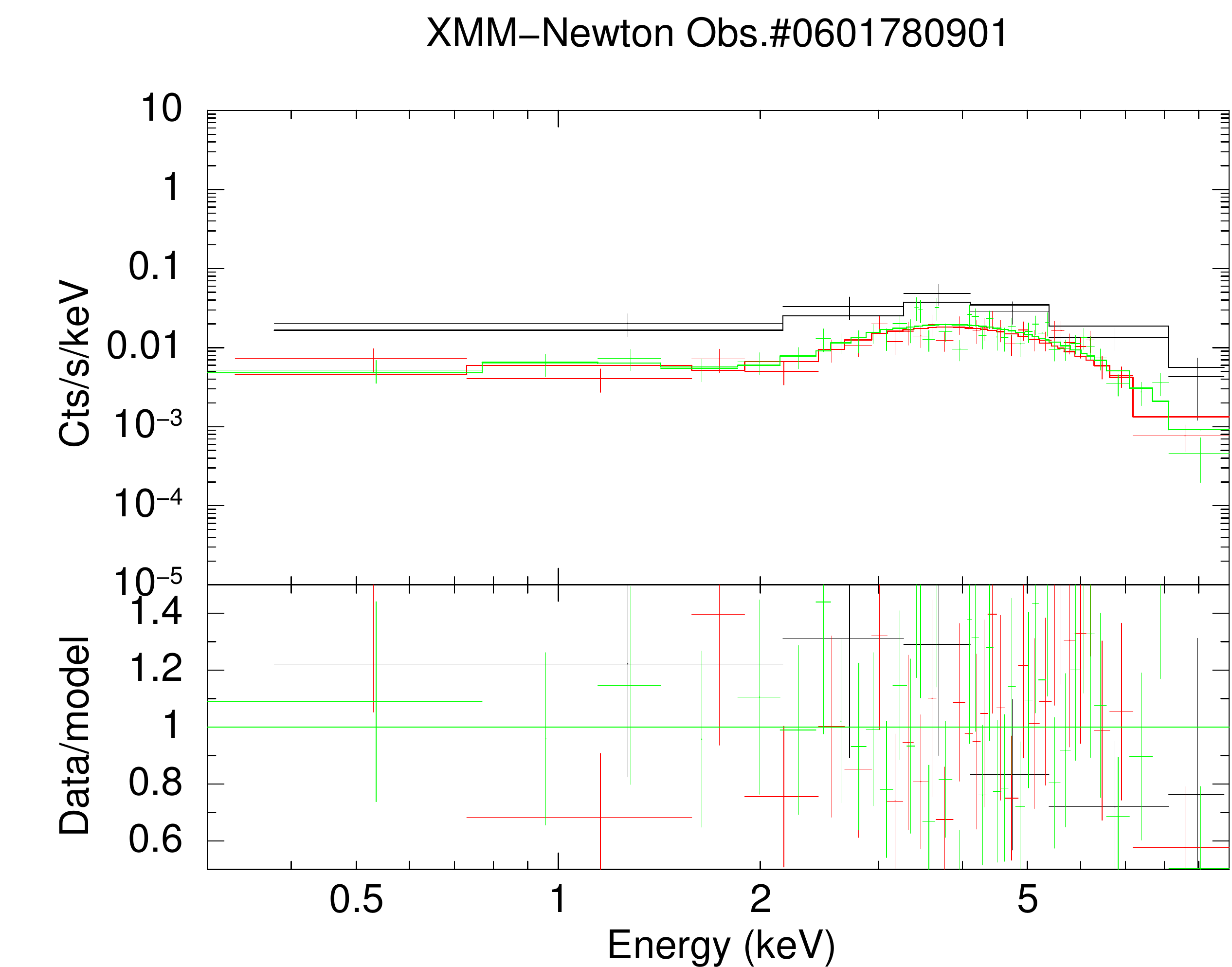}
   \includegraphics[width=0.225\textwidth, angle=-90]{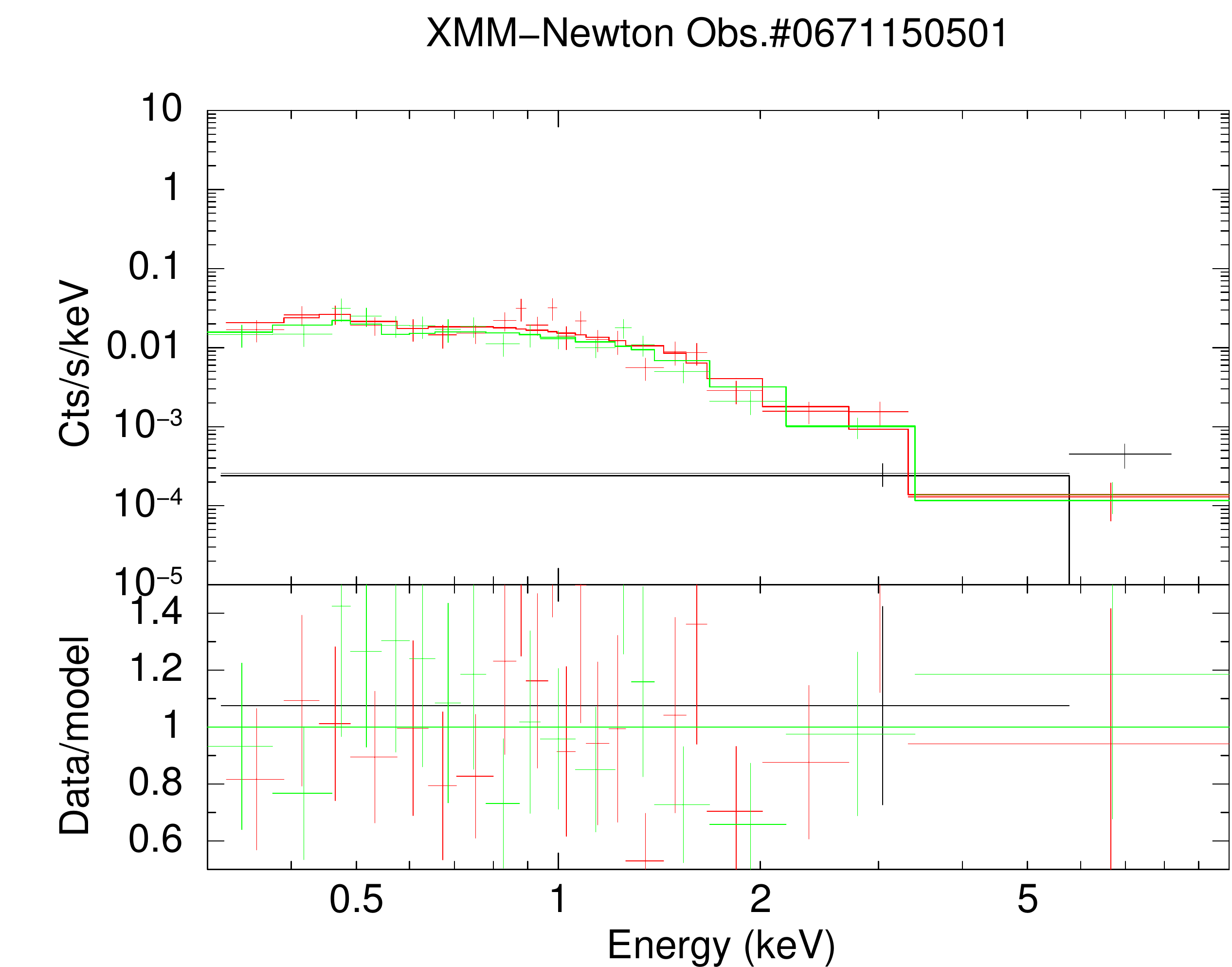}
   \includegraphics[width=0.225\textwidth, angle=-90]{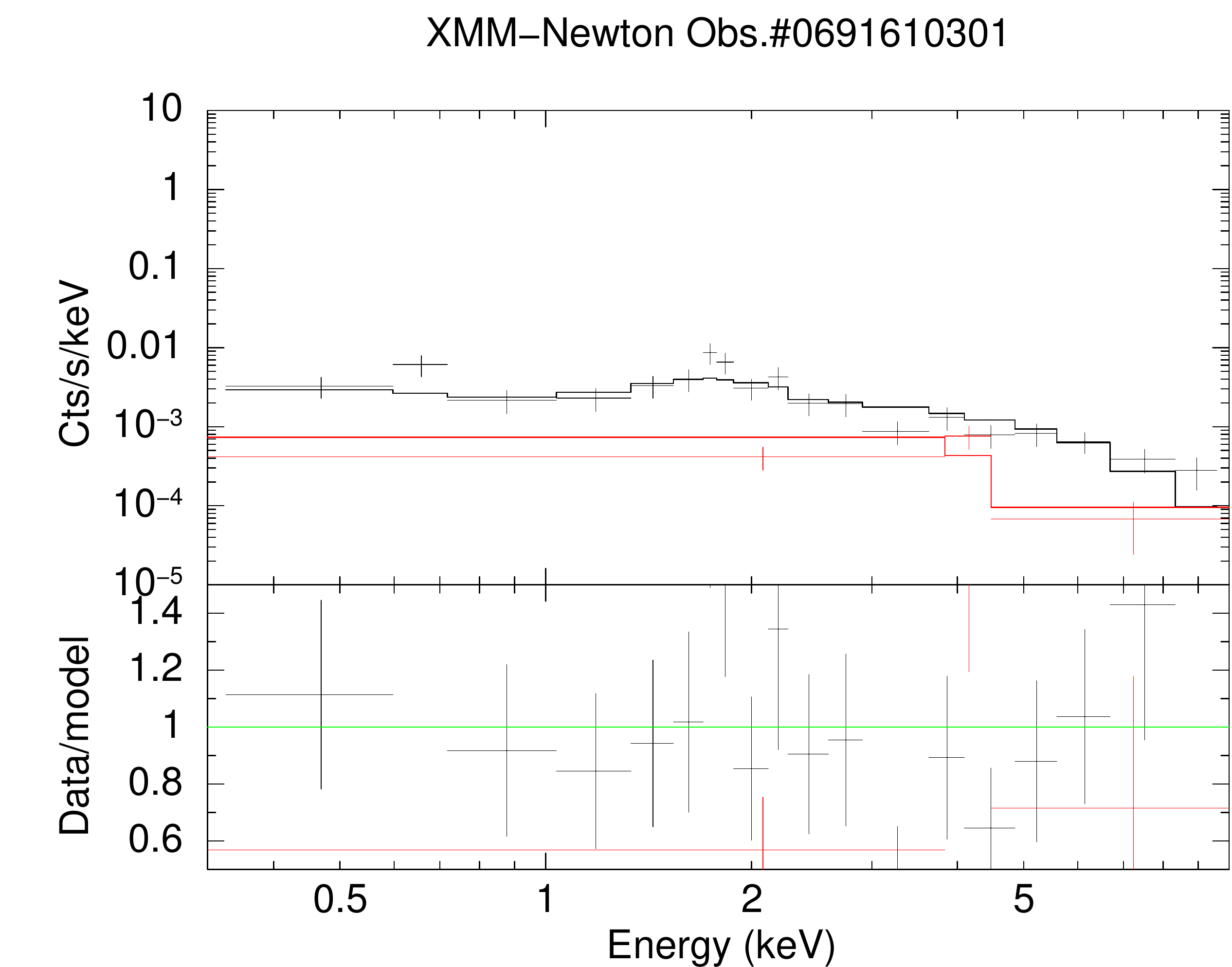}
   }
   \hbox{
   \includegraphics[width=0.225\textwidth, angle=-90]{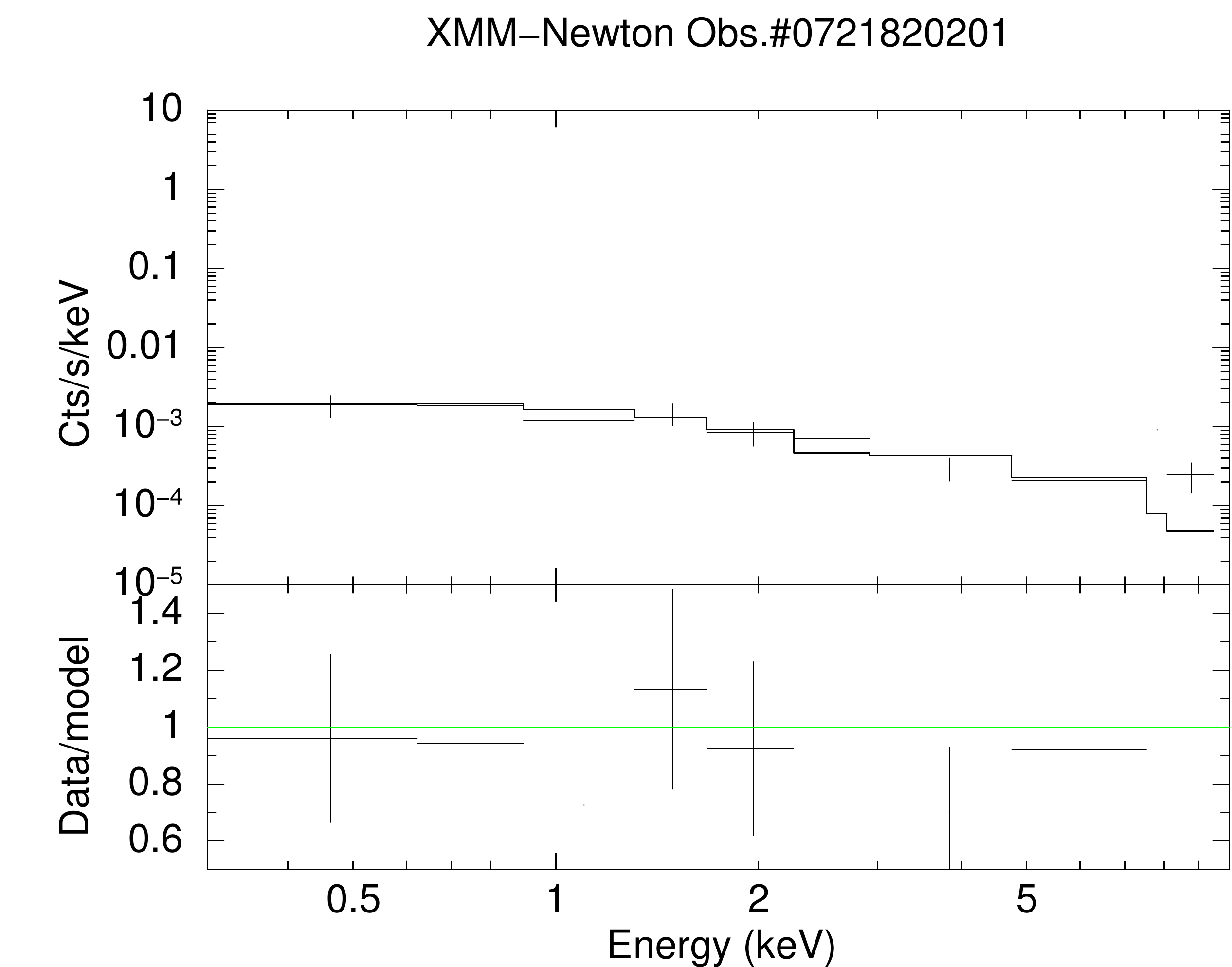}
   \includegraphics[width=0.225\textwidth, angle=-90]{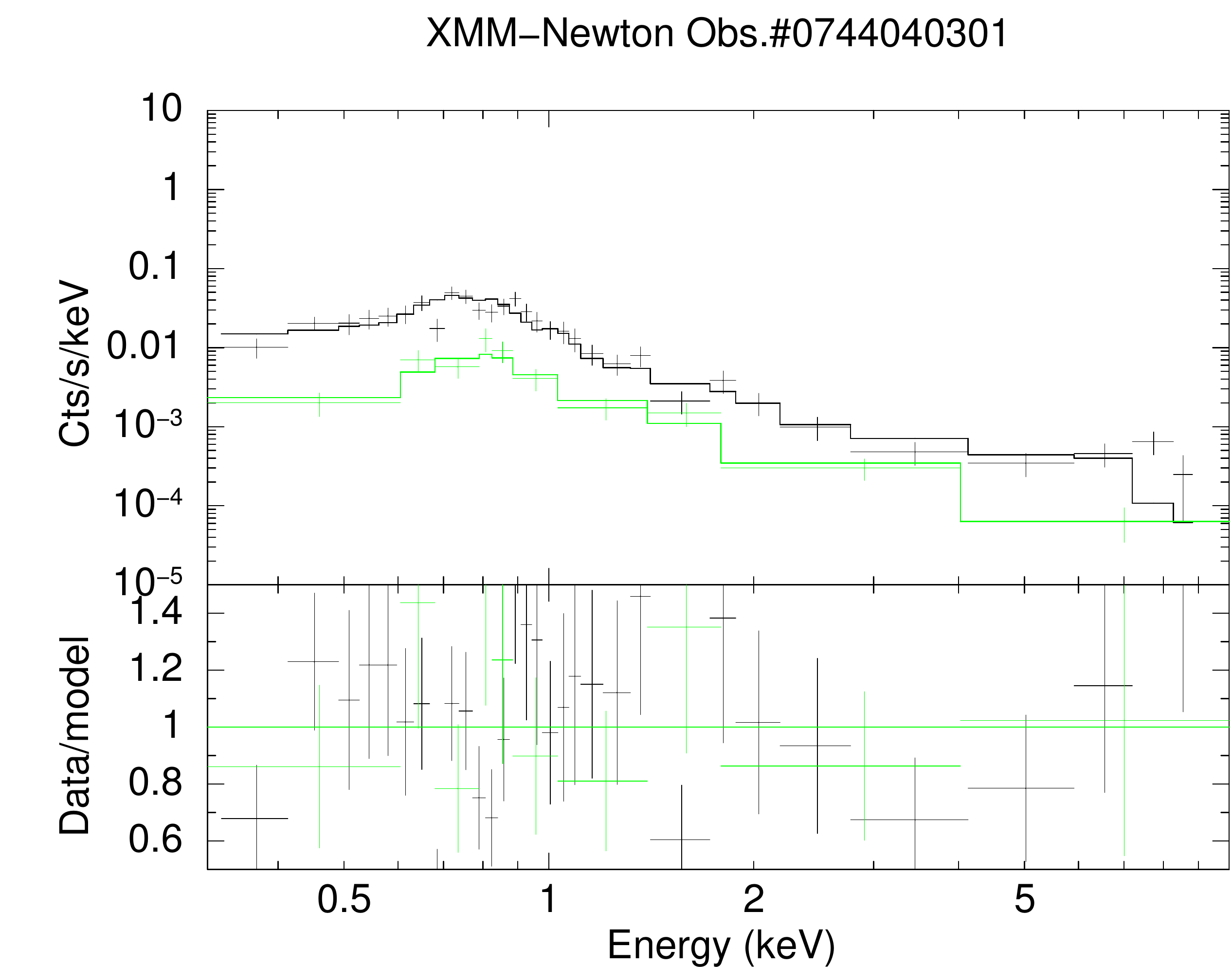}
   \includegraphics[width=0.225\textwidth, angle=-90]{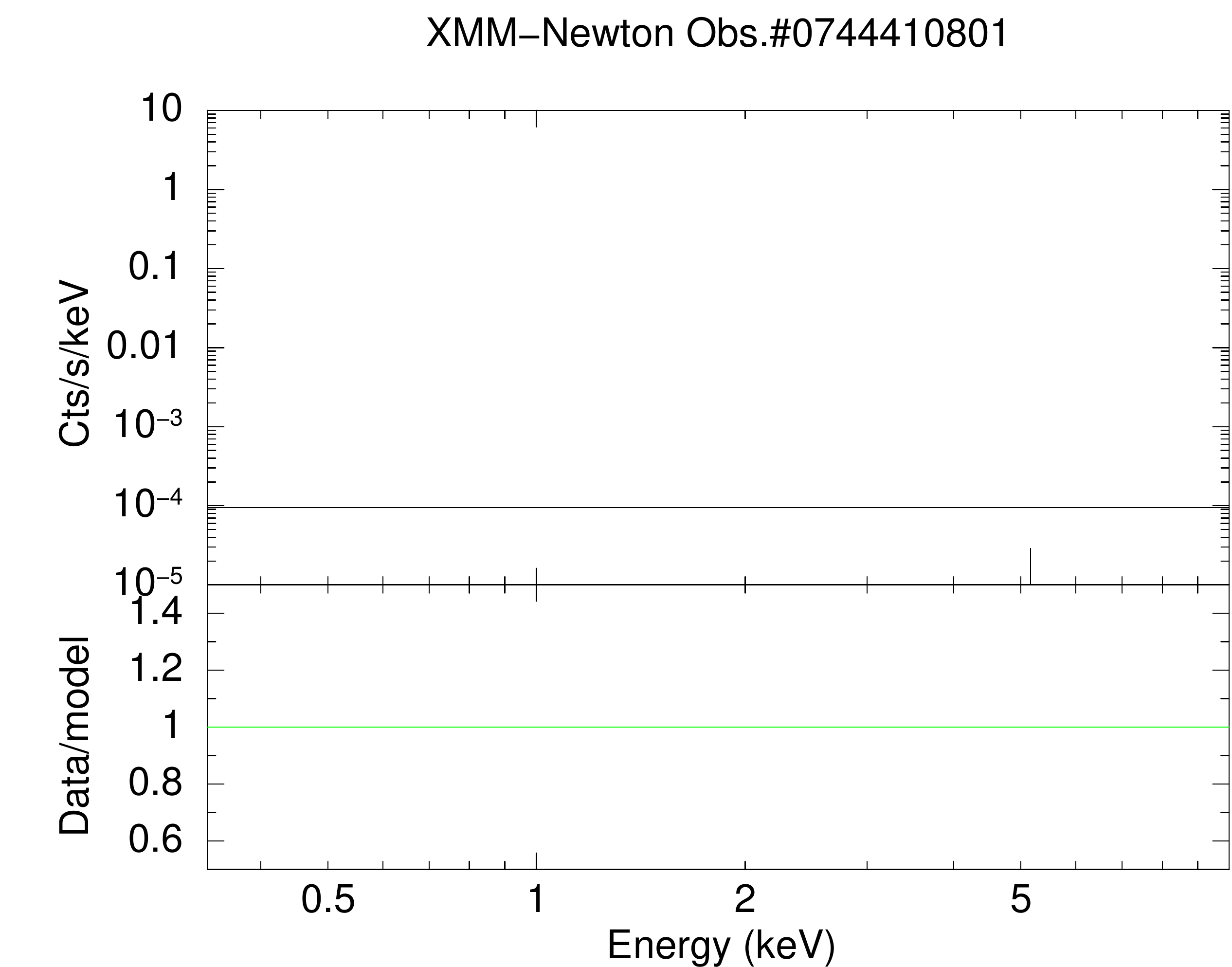}
   }
    \caption{{\it Upper panels}: EPIC spectra {\it crosses} and best-fit models {\it solid lines} for the 19 {\it bona fide} AGN spectra in our sample. {\it Lower panels}: residuals against the best-fit model in units of data/model ratio. {\it Black}: EPIC-pn; {\it red}: EPIC-MOS1; {\it green}: EPIC-MOS2. Each data point corresponds to a minimum signal-to-noise ratio $\ge$3.
}
    \label{fig:spectra1}
\end{figure*}
\begin{figure*}
    \centering
    \hbox{
   \includegraphics[width=0.225\textwidth, angle=-90]{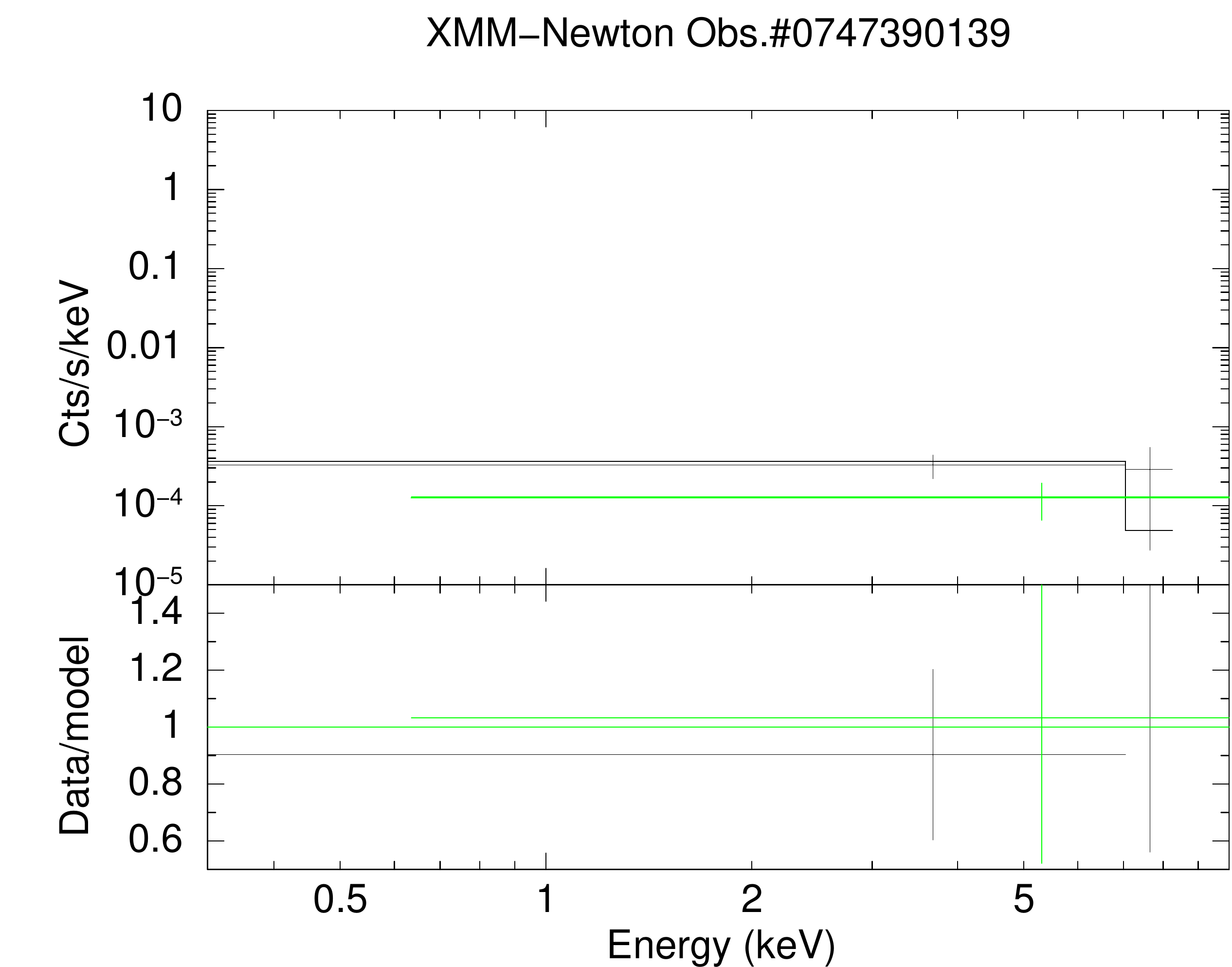}
   \includegraphics[width=0.225\textwidth, angle=-90]{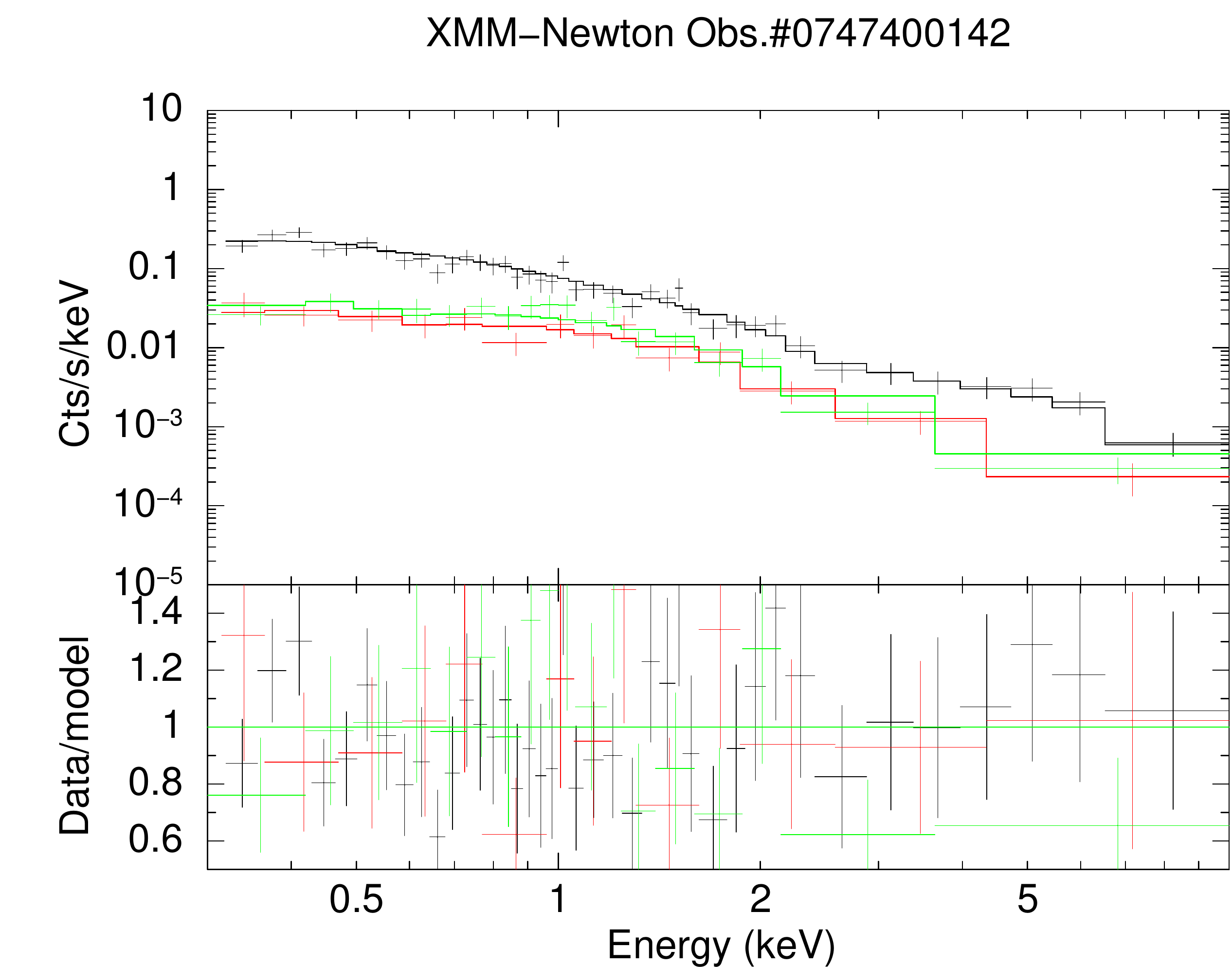}
   \includegraphics[width=0.225\textwidth, angle=-90]{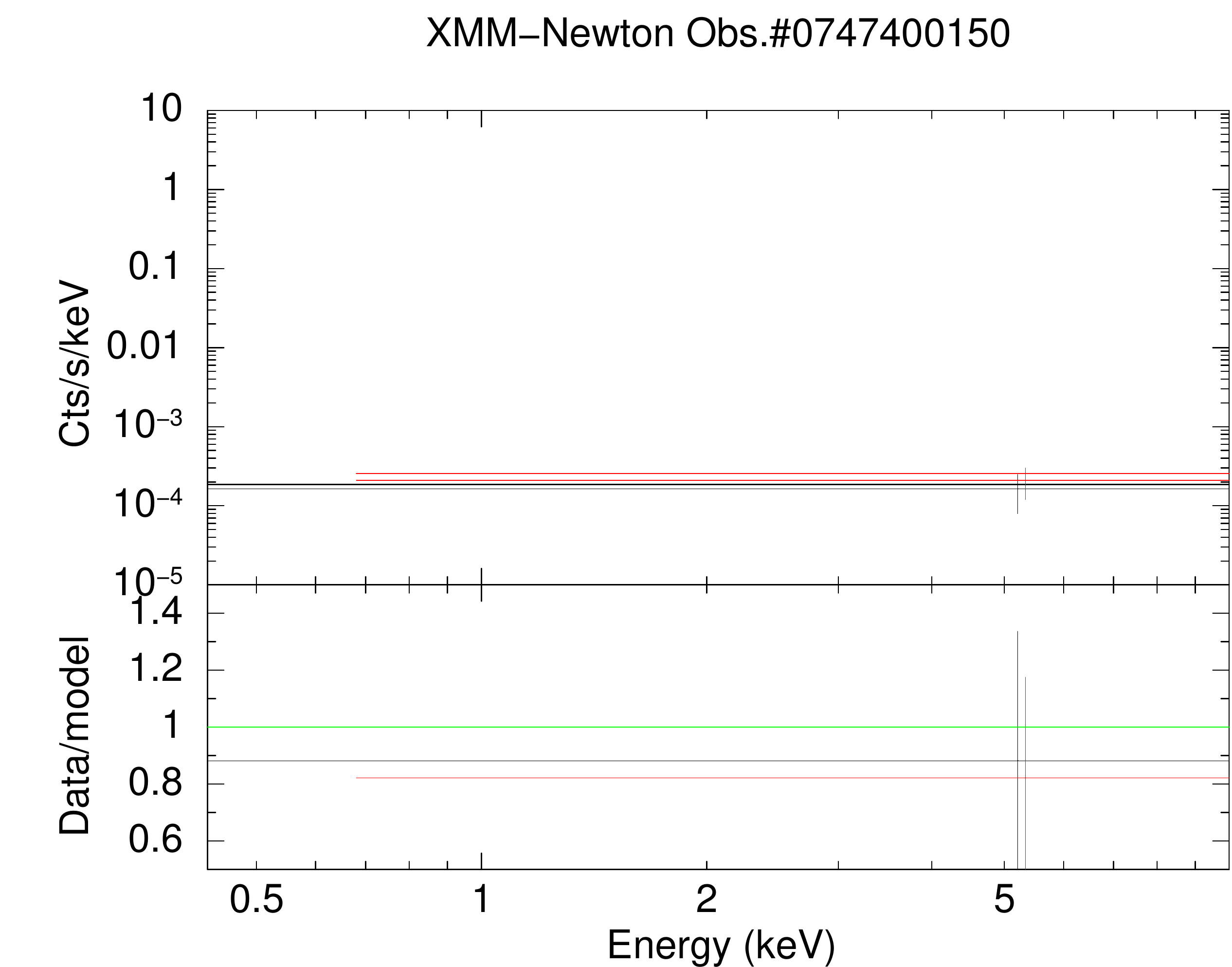}
   }
   \hbox{
   \includegraphics[width=0.225\textwidth, angle=-90]{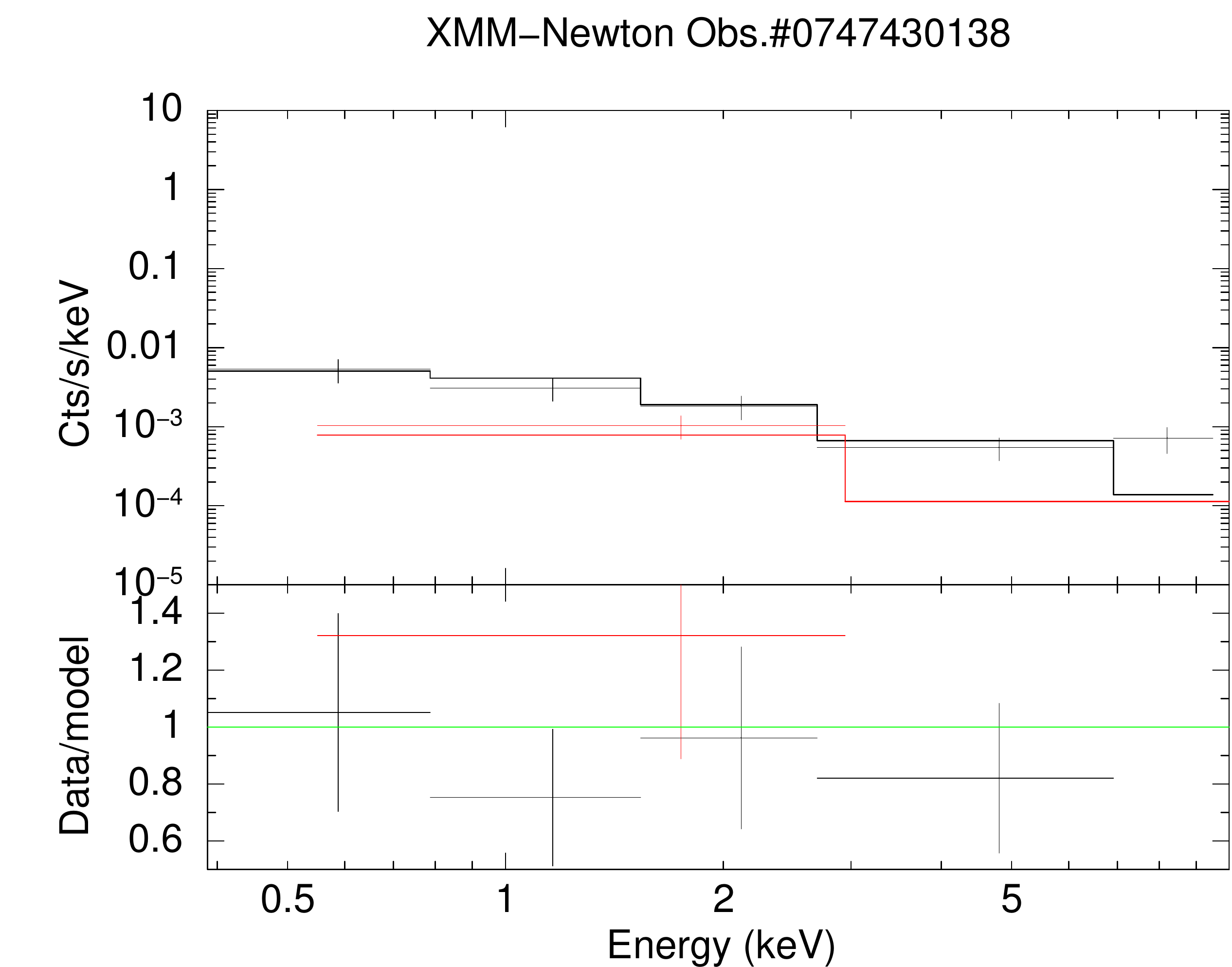}
   }
    \caption{Continuation of Fig.~\ref{fig:spectra1}.
}
    \label{fig:spectra2}
\end{figure*}

\section*{Data availability statement}
The high-level data underlying this article are extracted through standard processing from raw data
stored in public archives (SDSS and XMM-Newton), and will be shared on reasonable request to the corresponding author.

\bibliographystyle{mnras}
\bibliography{3XMM.bib}

\bsp	
\label{lastpage}
\end{document}